# Optical Coherence Elastography Measures Mechanical Tension in the Lens and Capsule

Xu Feng[1,4,†], Guo-Yang Li[1,3,†], Yuxuan Jiang[1,†], Owen Shortt-Nguyen[1], Seok-Hyun Yun[1,2,*]

[1]Harvard Medical School and Wellman Center for Photomedicine, Massachusetts General Hospital, 50 Blossom St., Boston, MA 02114, USA

[2]Harvard-MIT Division of Health Sciences and Technology, Cambridge, MA 02139, USA

[3]Currently with the Department of Mechanics and Engineering Science, College of Engineering, Peking University, Beijing 100871, China

[4]Currently with the Department of Bioengineering, University of Texas at Dallas, TX 75080, USA

[†]Equal contribution

*syun@hms.harvard.edu

**Abstract:** Lens tension is essential for accommodative vision but remains difficult to measure with precision. Here, we present an optical coherence elastography (OCE) technique that quantifies both tension and elastic modulus in the lens capsule and underlying tissue. This method derives mechanical parameters from surface wave dispersion across a critical frequency range of 1-30 kHz. Using isolated lenses from six-month-old pigs, we measured intrinsic anterior capsular tensions of 0-20 kPa and posterior capsular tensions of 40-50 kPa, induced by intra-lenticular pressure at the cortical surface. The mean shear moduli of anterior and posterior capsules were 630 kPa and 400 kPa, respectively, nearly 100-fold greater than that of the cortical tissues, where tensions were below 1 kPa. Biaxial zonular stretching (~4% strain) increased anterior capsular tension by 67 kPa, with a low uncertainty of only 2 kPa. This optical method holds significant promise for diagnosing and managing accommodative dysfunctions through lens mechanics assessment in clinical settings.

## 1. Introduction

The crystalline lens plays a critical role in human vision by enabling fine-tuning of focus through the process of accommodation. The gradient lenticular-shape structure, consisting of the nucleus, cortex and capsule, contributes approximately 30-35% of the eye's total refractive power. Focal adjustment is achieved by changes in lens shape, driven by tension from the surrounding zonular fibers attached to the capsule at the equator of the lens. These fibers are connected to the ciliary muscle, which receive neuronal signals from the visual cortex, forming a negative-feedback loop for focusing. Although some debate persist [1-3] regarding the precise mechanical interplay between the ciliary muscle, suspensory zonular fibers, and lens capsule, the Helmholtz accommodation theory, proposed in 1856, is widely accepted [4]. According to this theory, contraction of the ciliary muscle loosens the zonular fibers, releasing tension on the lens capsule. The relaxed lens adopts a more spherical and thicker shape, shortening its short focal length for near vision. Conversely, when viewing distant objects, the ciliary muscle relaxes, allowing the zonular fibers to stretch the lens. This tension flattens the lens [5, 6], increasing its focal length for far vision.

It is well established that the accommodative tuning range of focal length declines steadily with age. A significant loss of accommodative ability, known as presbyopia, typically occurs around the age of 45 and progresses to a complete loss by approximately age 65 [7, 8]. Presbyopia is primarily attributed to age-related stiffening and thickening of crystalline lens tissues [9, 10]. As the lens hardens, it becomes non-deformable, even under normal ciliary and capsular tension. While age-related changes in ciliary muscle function have also been proposed as a contributing factor [11, 12], multiple studies have demonstrated that ciliary muscles remain functional in older individuals, even after the complete loss of accommodation power [13]. The specific role of zonular and capsular tension in the development of presbyopia, however, remains unclear.

This incomplete understanding is likely due to the challenges of evaluating ciliary muscle function *in vivo* and accurately quantifying the tension applied to the lens tissue. The mechanical properties of lens tissues and their role in lens deformability have been extensively studied using various mechanical methods [14]. They include spinning tests [15-17], indentation [18-20], compression [21], bubble acoustics [22, 23], and atomic force microscopy [24, 25], Brillouin microscopy [26, 27], optical coherence elastography (OCE) [28-31] and ultrasound elastography [32]. Additionally, the mechanical properties of the lens capsule have been a focus of research, particularly in the context of cataract surgery. Investigations have employed methods such as inflation testing [33], stretching [34], stress-strain analysis [10, 35-37], nanoindentation [38] and atomic force microscopy [25, 39]. However, none of these techniques have been applied to measure tension in the intact lens. Developing a method to measure lens tension and thereby infer ciliary muscle function could provide a valuable clinical tool for advancing our understanding of the accommodative process and for characterizing accommodative dysfunction and presbyopia.

In this study, we present an OCE-based method for the non-destructive measurement of tensions in the lens, with a particular focus on capsular tension, offering high sensitivity at the kilopascal (kPa) scale. Leveraging a state-of-the-art wideband OCE system [40, 41], this approach enables the precise measurement of elastic wave velocities along the lens surface over a critical frequency range (1-30 kHz) encompassing the onset of leaky surface waves. By applying an advanced elastoacoustic modeling framework, we extracted key mechanical parameters, including capsular tension, and intra-lenticular pressure (ILP), and the elastic moduli of the capsule and underlying tissues. This method was further used to quantify capsular tension under lens stretching forces. Finally, the potential clinical applications of this technique are discussed.

## 2. Materials and methods

### 2.1. Porcine sample preparation and measurement procedures

Lenses were surgically extracted from six fresh porcine eyeballs (Research 87, Inc., Boylston, MA). OCE measurements were first conducted on both the anterior and posterior sides of five intact lenses. Subsequently, capsular tension was released by cutting the capsules around the equator dissecting micro scissors. OCE measurements were then repeated on both the anterior and posterior sides of the five relaxed lenses. For three of these lenses, the capsules were peeled off using precision tweezers, and OCE measurements were performed on the isolated lens capsules placed loosely on a holder without external tension. A home-built biaxial stretcher was used to stretch intact lenses along with the attached zonules in fresh porcine eye globes. For these preparations, the posterior segment and cornea were removed, the sclera was sectioned into four quadrants, and four hooks were attached to each quadrant to enable biaxial stretching. To estimate lens stretch, a 3 mm x 3 mm square ink mark was placed on the lens surface, and its deformation was recorded using a camera and analyzed with ImageJ software. All measurements were completed within 12 hours post-mortem to ensure sample integrity.

### 2.2. Optical coherence elastography (OCE)

A custom-built optical coherence tomography (OCT) system equipped with a swept laser source centered at 1300 nm was used [41, 42]. The system operated at an A-line rate of 43.2 kHz, with an optical power of less than 10 mW delivered to the sample. Mechanical stimulations were applied using a custom contact probe [42] consisting of a piezoelectric transducer (PZT) and a 3D-printed, 2-mm plastic tip. A gentle contact force of approximately 0.01 N was applied, ensuring minimal deformation of the lens surface. Figure 1A shows the experimental setup, where a contact PZT probe, vibrating at a given frequency, generates elastic waves, and the vertical displacement is detected by phase-sensitive OCT.

Pure tone stimuli were used across a frequency range of 1-30 kHz. For each frequency, 172 A-lines were collected (M-scan) while the OCT beam was maintained fixed at a location on the sample. The M scan was repeated at 96 transverse positions on the lens. The scan length varied inversely with frequency, ranging from 3 mm at 1-6 kHz to 1 mm at 20-30 kHz, corresponding to 2-3 times the wavelength at each frequency. These scan lengths were chosen to capture exponentially decaying wave profiles efficiently. Displacement profiles over time ($t$) were extracted at each scan position and Fourier-transformed to convert the data into the frequency ($f$) domain. The phase and amplitude of the mechanical vibration were filtered using a lock-in bandwidth of ~250 Hz around the excitation frequency. Two-dimensional cross-sectional displacement images were generated by performing these steps for all scan positions. To calculate phase velocities, one-dimensional complex displacement profiles along the sample surface were extracted and Fourier-transformed to obtain displacement spectra in the spatial frequency (wavenumber) domain. The wavenumber corresponding to the peak amplitude was identified as $k$, then the phase velocity $v$ was calculated using the equation $v = 2\pi f/k$. A dispersion curve was constructed from the phase velocities corresponding to each stimulus frequency. More details of this data processing can be found in our previous work [41, 43].

### 2.3. Bilayer guided wave model under tension

We modeled the lens using a tension-induced bilayer framework comprising a stiff top layer with thickness $h$ and shear modulus $\mu_1$, and a semi-infinite soft substrate with shear modulus $\mu_2$ (Fig. 1B). The stiff top layer represents the capsule while the soft substrate corresponds to the cortical tissue. The lens nucleus was excluded from the model, as its impact on wave dispersion was negligible at frequencies below 1 kHz. Both layers were assumed to be elastic, isotropic, and incompressible. The biaxial stresses in the top layer and substrate are denoted as $\sigma_1$ and $\sigma_2$, respectively. Guided waves in this bilayer structure are subject to stress-free boundary conditions at the air-facing surface and continuity conditions at the interface between the layers. Using incremental dynamics theory [44, 45], the secular equation governing elastic guide wave propagation in the bilayer structure is derived (see details in Section 3.1 and Supplementary Note 1):

$$\det(\mathbf{M}_{6\times 6}) = 0, \quad (1)$$

where $\mathbf{M}_{6\times6}$ is a matrix with components given in Appendix. The relationship between wave angular frequency $\omega$ $(= 2\pi f)$ and wavenumber $k$ is determined by Eq. (1). The phase velocity $v$ is then obtained as $v = \omega/\mathrm{Re}(k)$.

## 2.4. Inverse method to extract mechanical parameters

Mechanical parameters were extracted from measured wave velocity dispersion curves using analytical solutions derived from bilayer model. In the stress-free state, the dispersion is described by $v = v(h, \mu_1, \mu_2; f)$, where $f$ represents the stimulus frequency, and $h$, $\mu_1$, and $\mu_2$ are input variables. The capsular thickness $(h)$ was directly obtained from OCT images. The substrate modulus $(\mu_2)$ was calculated using $\mu_2 = \rho {v_{t2}}^2$, where $v_{t2}$ was estimated from wave velocities measured at low frequencies (1-2 kHz). The only remaining variable, $\mu_1$ (the capsule modulus), was determined by fitting the experimental dispersion curves in the stress-free state. Curve fitting was performed by minimizing the root-mean-square-error (RMSE) as the objective function, with the best fit found by identifying the minimum error. In the pre-stressed state, the dispersion is described by $v = v(h, \mu_1, \mu_2, \sigma_1, \sigma_2; f)$, where $h$, $\mu_1$, and $\mu_2$ has already been determined from the stress-free state. The remaining variables, the stresses $\sigma_1$ and $\sigma_2$, were extracted by fitting the dispersion curves measured in the pre-stressed state. Similar to the stress-free state, the best fit was identified by minimizing the RMSE, defined as $\mathrm{RMSE} = \sqrt{\sum_{i=1}^{n}\left(v_i^{(\mathrm{exp})} - v_i^{(\mathrm{model})}\right)^2/n}$, where $v_i^{(\mathrm{exp})}$ and $v_i^{(\mathrm{model})}$ denote the experimentally measured and model-predicted phase velocities, respectively, and *n* is the number of data points (*n* = 12) within the frequency range of 1 - 30 kHz.

## 2.5. Finite-element numerical modeling

Finite element analysis (FEA) was performed using Abaqus/CAE 6.14 (Dassault Systèmes, USA). An axisymmetric bilayer ellipsoid-shape model was developed to simulate the lens capsule and cortex. To simulate tension from the zonular fibers, the lens model was initially subjected to a quasi-static stretch applied at multiple points on the side of the lens. Subsequently, a fixed-frequency excitation was applied to the top of the posterior capsule to model elastic wave propagation and dispersion. The model parameters were chosen to closely match our experimental data. The model was discretized using approximately 110,000 axisymmetric solid elements (CAX8RH). Simulation convergence was carefully verified by comparing results with those obtained using a finer mesh and smaller time steps. The phase velocity was extracted from the spatiotemporal displacement field on the surface of the capsule. A Fourier transform was applied to determine the wavenumber ($k$) at each frequency ($f$). The phase velocity ($v$) was calculated using $v = 2\pi f/k$.

## 2.6. Statistical analysis

Phase velocity data for the ex vivo porcine lenses are presented as mean ± standard deviation (S.D.). Paired Student's t-test was used to compare phase velocities between the intact anterior and posterior lens, as well as between intact and incised states for both the anterior and posterior lenses. Comparisons were also made for the mechanical parameters $\mu_1$, $\mu_2$, $\sigma_1$, $\sigma_2$ between the anterior and posterior sides. A *p*-value < 0.05 was considered statistically significance. Results with $0.01 \leq p \leq 0.05$ are indicated with one asterisk (*); $0.001 \leq p < 0.01$ with two asterisks (**); and $p < 0.001$ with three asterisks (***).

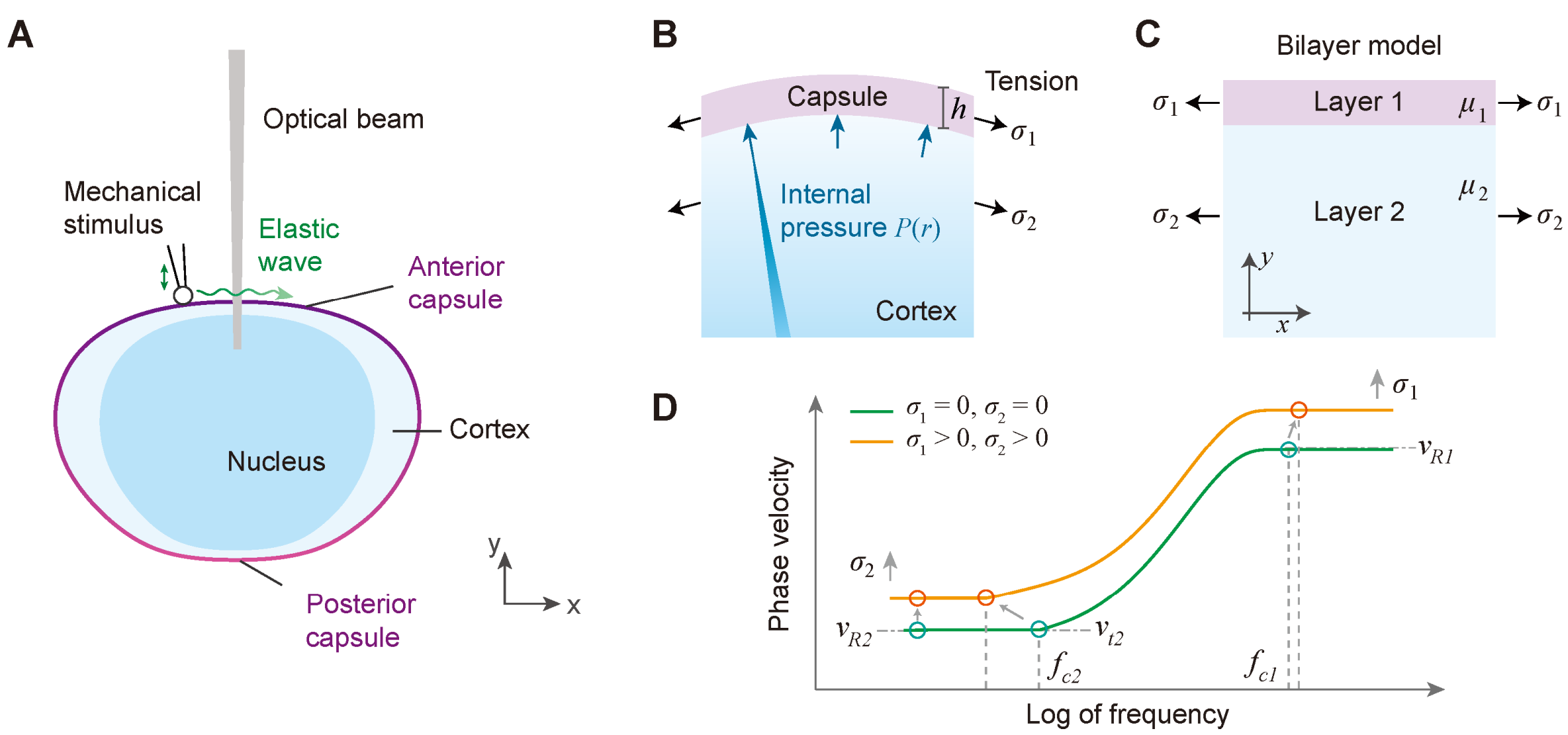

**Fig. 1**. The lens and elastic waves under tension. (**A**) Schematic of the experimental setup to measure acoustic wave profiles using optical coherence elastography. (**B**) Schematic of the lens surface, illustrating transverse tensions and the radial gradient of intra-lenticular pressure (blue arrows). (**C**) Equivalent bilayer model comprising a top layer (layer 1) and a semi-infinite substrate (layer 2). Shear moduli are denoted as $\mu_1$ and $\mu_2$ ($< \mu_1$). (**D**) Phase velocity of the guided wave as a function of frequency. At low frequencies, the phase velocity approaches the Rayleigh wave velocity of the substrate (lens tissue), $v_{R2}$. As the frequency increases, it reaches the shear wave velocity of the substrate, $v_{t2}$, at the critical frequency $f_{c2}$. At high frequencies beyond another critical point, $f_{c1}$, the velocity plateaus, approaching the Rayleigh wave velocity of the top layer (capsule), $v_{R1}$. Dots mark the initial and key transition points along the dispersion curve. Increased tension raises the overall velocities, decrease $f_{c2}$, and increases $f_{c1}$.

## 3. Results

### 3.1. A bilayer mechanical model under tension

Figure 1B illustrates the lens model under tension. The ILP is the highest at the nucleus center and decreases toward the lens periphery, varying as a function of the radial position $r$. For isolated lenses without external force, the capsular tension $\sigma_1$ arises from the internal pressure, $P(r)$; under force equilibrium [46],

$$\sigma_1 = \frac{P(R)R}{2h}, \tag{2}$$

where $R$ is the radius of curvature of the lens surface, and $h$ is the thickness of the capsule. The pressure gradient also induces deviatoric lateral stress in the lens tissue. By force balance (See Supplementary Note 2),

$$\sigma_2(r) = \frac{r}{2}\frac{dP}{dr}. \tag{3}$$

When lenses are subjected to external forces, such as tension from zonular fibers, the capsular and cortical tensions are modified as follows:

$$\sigma_1 = \sigma_1^{ext} + \frac{P(R)R}{2h} \tag{4}$$

$$\sigma_2(r) = \sigma_2^{ext} + \frac{r}{2}\frac{dP}{dr}, \tag{5}$$

where $\sigma_1^{ext}$ and $\sigma_2^{ext}$ represent the tensions induced by the external force in the capsule and cortex, respectively.

We simplified the lens model as a flat bilayer system (Fig. 1C). The top layer represents the capsule, with shear modulus $\mu_1$, uniform biaxial stress $\sigma_1$, and wall thickness $h$. The semi-infinite substrate represents the cortex, characterized by a uniform shear modulus $\mu_2$ ($< \mu_1$) and uniform biaxial stress $\sigma_2 \approx \sigma_2(r)|_{r=R}$ . The lens nucleus was excluded from the model, as its impact on wave dispersion was negligible at frequencies below 1 kHz. Both layers were assumed to be elastic, isotropic, and incompressible. For each layer ($i$ = 1, 2), incremental parameters $\alpha_i$, $\beta_i$ and $\gamma_i$ were derived from the fourth-order Eulerian elasticity tensor $\boldsymbol{\mathcal{A}}^{(i)}$as follows [45]: $\alpha_i = \mathcal{A}_{xyxy}^{(i)}$, $2\beta_i = \mathcal{A}_{xxxx}^{(i)} + \mathcal{A}_{yyyy}^{(i)} - 2\mathcal{A}_{xxyy}^{(i)} - 2\mathcal{A}_{xyyx}^{(i)}$, $\gamma_i = \mathcal{A}_{yxyx}^{(i)}$. Under small-strain biaxial stretching, the relationships are (see Supplementary Note 1):

$$\alpha_i = \mu_i + \frac{1}{3}\sigma_i,\ \beta_i = \mu_i - \frac{1}{6}\sigma_i,\ \gamma_i = \mu_i - \frac{2}{3}\sigma_i. \tag{6}$$

Figure 1D illustrate the solutions of the secular equation. At low frequency limit, the wave speed approaches a plateau at $v_{R2}$, and at high frequencies, it asymptotically reaches $v_{R1}$, where (see Supplementary Note 3):

$$\rho_i {v_{Ri}}^2 = 0.913\mu_i + 0.392\sigma_i, \tag{7}$$

where $\rho_i$ are the mass density of each layer. Note that $v_{R1}$ and $v_{R2}$ are similar to the bulk wave speeds $v_{ti}$ and exhibit slightly higher sensitivity to tension. The bulk shear wave speed in an infinitely large layer is given by

$$\rho_i {v_{ti}}^2 = \mu_i + \sigma_i/3, \tag{8}$$

The transition between these regimes begins at a critical frequency, $f_{c2}$, where the wave dispersion curve bifurcates into leaky surface-guided and bulk-wave-like branches. The critical frequency is given by

$$f_{c2} = C_1 \left(\frac{\mu_2}{\mu_1}\right)^{C_2} \frac{v_{t2}}{h}, \tag{9}$$

where $C_1 \approx 73.7\left(\frac{\sigma_1}{\mu_1}\right)^3 - 26.8\left(\frac{\sigma_1}{\mu_1}\right)^2 + 2.4\left(\frac{\sigma_1}{\mu_1}\right) + 0.16$ and $C_2 \approx 1 - 0.5\exp\left(-30\frac{\sigma_1}{\mu_1}\right)$ (see Supplementary Note 3). In the absence of tension ($\sigma_1$ = 0), $f_{c2}$ simplifies to the result previously derived for tension-free bilayers [42]. Within the transition region, the guided wave velocity increases with frequency. The high-frequency plateau is reached at another critical frequency, $f_{c1} \approx v_{t1}/h$.

This formalism provides a powerful analytical tool for determining the key mechanical parameters $\sigma_1$, $\sigma_2$, $\mu_1$, and $\mu_2$ from experimentally obtained wide-frequency velocity data. For typical accommodative human lenses, $f_{c2}$ is expected to range between 100 Hz and 10 kHz, and $f_{c1}$ would be close to 1 MHz. These frequencies fall within the capabilities of state-of-the-art OCE systems [40, 41]. However, accurately measuring wave velocities becomes increasingly difficult at higher frequencies, typically beyond 50 kHz, due to reduced wave displacement. Notably, dispersion data over a substantial frequency range, such as 1-30 kHz, can still reveal sufficient features to determine the key parameters through curve fitting.

### 3.2. Wideband OCE measurements on isolated intact lenses

We performed OCE measurements on freshly isolated lenses from six-month-old pigs, measuring both the anterior and posterior sides (Fig. 2A). For the posterior lens measurements, the sample was flipped. OCT structural images (Fig. 2B) show that the anterior capsule is thicker than the posterior capsule. Figure 2C presents OCE wave maps recorded at 2 kHz and 10 kHz, where wavelength decreases with increasing frequency on both surfaces. At 10 kHz, multiple wave cycles are observed. Previously, we have observed stronger attenuation of Rayleigh waves at 10 kHz in the cornea and skin [41-43], but here, the results indicate that the lens tissue exhibits a dominance of elasticity over viscosity. The velocity data revealed significant differences between the anterior and posterior lens surfaces. At low frequencies (1-2 kHz), phase velocities were only modestly different (Fig. 2D, 2 kHz), indicating similar cortical stiffness on both sides. However, at higher frequencies (> 4 kHz), more pronounced differences emerged (Fig. 2D, 10 kHz).

Figure 2E shows the wave velocities measured along the lens surface across the frequency range of 1 to 30 kHz. They exhibit a strong frequency dependency. While the viscoelastic properties of tissue contribute to frequency-dependent wave speeds, such pronounced dependency typically suggests a stiff surface overlying a softer interior, similar to tissues such as skin [42]. At low frequencies, where the wavelength far exceeds the thickness of the capsule, the capsule has minimal impact on wave velocity, which is mainly determined by the softer lens tissue. As the frequency increases, the wave's elastic energy becomes progressively concentrated within the stiffer capsule, resulting in higher velocities. Using the bi-layer tensioned lens model, the best fit to the experimental data was achieved with the following parameters: for the anterior lens, $h$ = 55 μm, $\sigma_1$ = 12 kPa, $\mu_1$ = 350 kPa, $\sigma_2$ = 180 Pa, and $\mu_2$ = 5.6 kPa; and for the posterior lens, $h$ = 15 μm, $\sigma_1$ = 41 kPa, $\mu_1$ = 400 kPa, $\sigma_2$ = 390 Pa, and $\mu_2$ = 2.4 kPa.

OCE measurements were performed on five different 6-month-old porcine lenses. Figure 2F-2I show the fitted parameters (mean ± S.D.) obtained using the bi-layer model. For the anterior lens, we found $\sigma_1$ = 8 ± 8 kPa, $\sigma_2$ = 150 ± 30 Pa, $\mu_1$ = 630 ± 273 kPa, and $\mu_2$ = 6 ± 0.3 kPa. For the posterior sides, the parameters were $\sigma_1$ = 44 ± 4 kPa, $\sigma_2$ = 800 ± 600 Pa, $\mu_1$ = 440 ± 107 kPa, and $\mu_2$ = 2.3 ± 0.7 kPa.

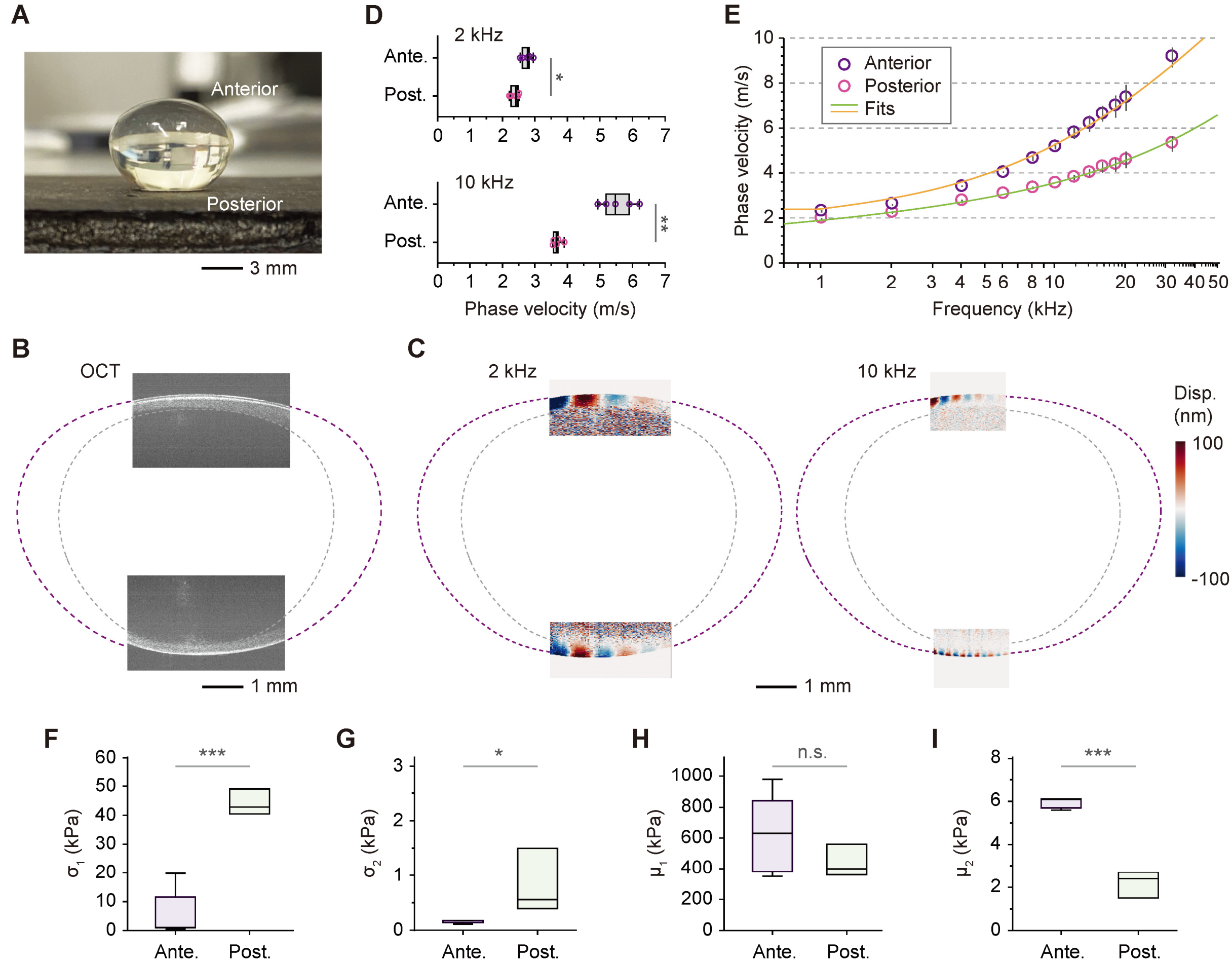

**Fig. 2.** OCE measurement on intact porcine lenses. (**A**) Photography of a 6-month-old porcine lens. (**B**) Cross-sectional OCT image of the lens. (**C**) OCE wave displacement profiles recorded at 2 kHz and 10 kHz. Dashed outlines denote the external surface, and dotted lines mark the interface between the cortex and nucleus. (**D**) Measured phase velocities at 2 kHz (top) and 10 kHz (below) in the anterior and posterior capsules. $^*p = 0.017$; $^{**}p = 0.0015$, determined by Student's t-test (n = 5). (**E**) Wave velocities measured on the anterior and posterior surfaces, presented as mean +/- S.D. (n = 3 measurements at different locations on one sample). Circles: experiment data; lines: model fit curves. (**F**) Statistical results of capsular tension. (**G**) Statistical results of cortical tension. (**H**) Statistical results of capsular shear modulus. (I) Statistical results of cortical shear modulus. $^*p < 0.05$; $^{***}p < 0.001$, n.s., not significant (n = 5).

### 3.3. Quantification of intrinsic capsular tension

Six-month-old porcine lenses exhibited high elasticity during finger palpation. When a small incision was made on one side of the capsule, lens tissue visibly bulged through the opening (Fig. 3A). OCE measurements were performed on lenses before and after a complete circumferential incision of the capsule around the equator (Fig. 3B). Following the incision, the posterior capsule exhibited noticeable shrinkage (~ 3%) as observed visually, whereas the anterior capsule showed less pronounced shrinkage. This indicates that the ILP-induced capsular tension in intact lenses is substantially greater in the posterior capsule than in the anterior.

Consistent with this observation, the anterior lens surface displayed only minor changes in wave profiles (Fig. 3C), and five samples showed moderate changes in surface wave velocities at 6 kHz ($p$ = 0.05; Fig. 3D). Velocity dispersion analysis revealed moderate differences before and after incision (Fig. 3E). In contrast, the posterior surface showed significant changes in wave profiles (Fig. 3F), velocities ($p$ = 0.003, n = 5; Fig. 3G), and dispersion curves (Fig. 3H) following the release of capsular tension. Notably, wave attenuation on the posterior surface increased markedly, indicating a transition from predominantly elastic to viscoelastic behavior (Fig. 3F). Model fitting based on the data in Fig. 3E and 3H yielded the following parameters: for the anterior lens, $\sigma_1$ = 20 kPa, $\sigma_2$ = 200 Pa, $h$ = 55 μm, $\mu_1$ = 830 kPa, and $\mu_2$ = 5.7 kPa; and for the posterior lens, $\sigma_1$ = 41 kPa, $\sigma_2$ = 400 Pa, $h$ = 15 μm, $\mu_1$ = 400 kPa, and $\mu_2$ = 2.3 kPa. Post-incision dispersion curves were accurately captured by the model assuming zero tension ($\sigma_1$ = $\sigma_2$ = 0). Together with the physical observation in Fig. 3A, these results strongly suggest that ILP is the main driver of capsular tension in isolated lenses.

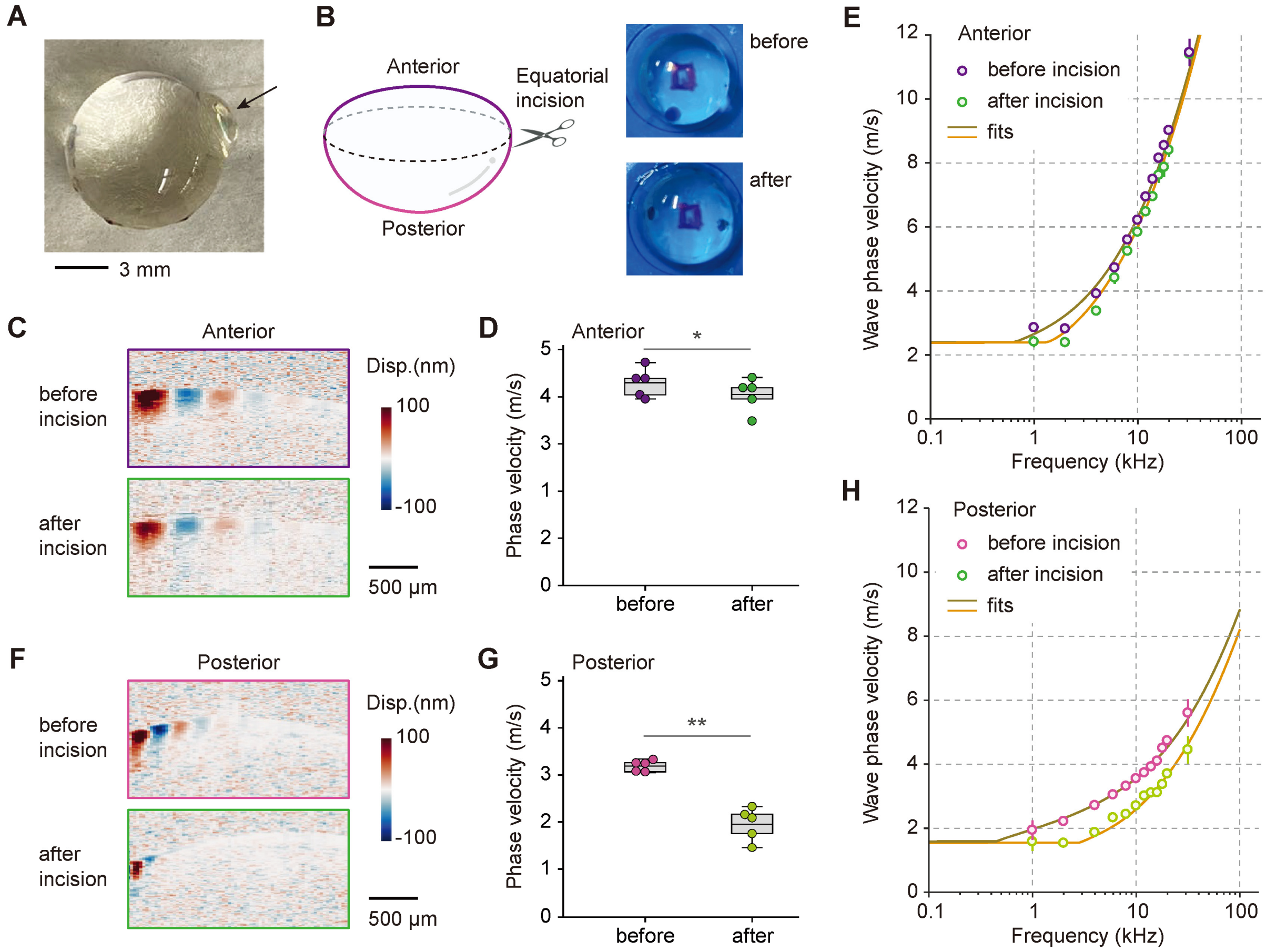

**Fig. 3**. The measurement of ILP-induced capsular tension. (**A**) Photograph showing lens tissue bulging through a small opening in the capsule. (**B**) Schematic of the circumferential incision made around the lens capsule. (**C**) Wave profiles on the anterior lens surface before and after incision at 6 kHz. (**D**) Phase velocities at 6 kHz on the anterior surface across five samples (*$p$ = 0.05). (**E**) Velocity dispersion curves for the anterior lens before and after incision. (**F**) Wave profiles on the posterior lens surface before and after incision at 6 kHz. (**G**) Phase velocities at 6 kHz on the posterior surface across five samples (**$p$ = 0.003). (**H**) Velocity dispersion curves for the posterior lens before and after incision. Error bars in (E), and (H) represent standard deviation from three measurements at different locations on a single sample.

Capsules were further isolated from the lenses (Fig. 4A) and placed loosely on a holder without external tension. OCE measurements demonstrated elastic waves propagation along the free-hanging anterior and posterior capsules (Fig. 4B). Wave velocity dispersion curves were obtained (Fig. 4C) and fitted using a tension-free single-layer Lamb-wave model [47, 48]. The best-fit parameters obtained from three specimens from three different lenses were $\mu_1$ = 417 ± 50 kPa for the anterior capsule and $\mu_1$ = 350 ± 40 kPa for the posterior capsule. These values match those previously derived from the intact lens using the bilayer model under tension, within the uncertainty of the fitting.

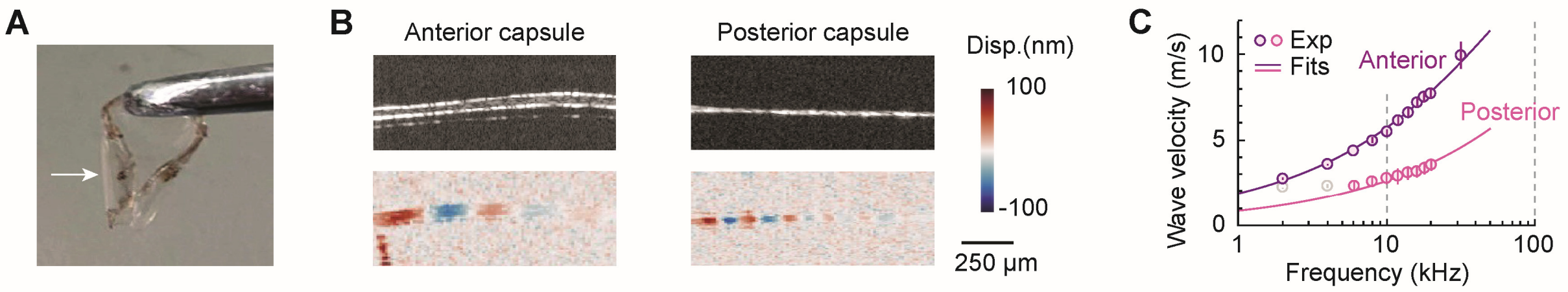

**Fig. 4**. OCE measurements on isolated lens capsules. (**A**) Photograph of an isolated anterior capsule. (**B**) Cross-sectional image (top) and wave profile at 12 kHz (bottom) of an isolated anterior capsule (left) and an isolated posterior capsule (right). (**C**) Wave velocity dispersion curves for the isolated capsules. Error bars represent standard deviation from 3 measurements at different locations of one sample. Dots: experimental data; Lines: fits based on the Lamb wave model.

### 3.4. A finite element model of the lens and zonular stretching

Finite element analysis (FEA) was performed using an axisymmetric lens model that included zonular ciliary fibers. Figure 5A shows the stress distribution in the lens under a radial stretch of 0.5 mm (5 mm in radius, ~10% stretch). The anterior capsule tension was approximately 70 kPa, while cortical lateral stress ranged from 700 Pa on the anterior side to 2 kPa at the equator. These results are consistent with previous reported values in the literature [49]. Figure 5B displays the displacement profile of elastic waves generated by a 10-kHz stimulus applied to the anterior capsule. The surface wave velocity was extracted by performing a Fourier transform of the displacement profile along the capsule. Figure 5C shows wave velocities (circles) measured under three different anterior capsular tension conditions: 0, 60, and 100 kPa, induced by varying the ciliary stretch. The dispersion curves (solid lines) calculated using the bilayer model show excellent agreement with the FEA results. Slight deviations near low critical frequencies are attributed to the inhomogeneous stress distribution in the cortex in the FEA model, compared to the assumption of uniform stress in the bilayer model. Figure 5D presents a super-linear relationship between the anterior capsular tension versus applied stretch. Figure 5E plots the change in lens thickness versus applied stretch, showing lens thickness reduction with stretch [50]. Figure 5F plots the stretching force as a function of stretch distance [49-51].

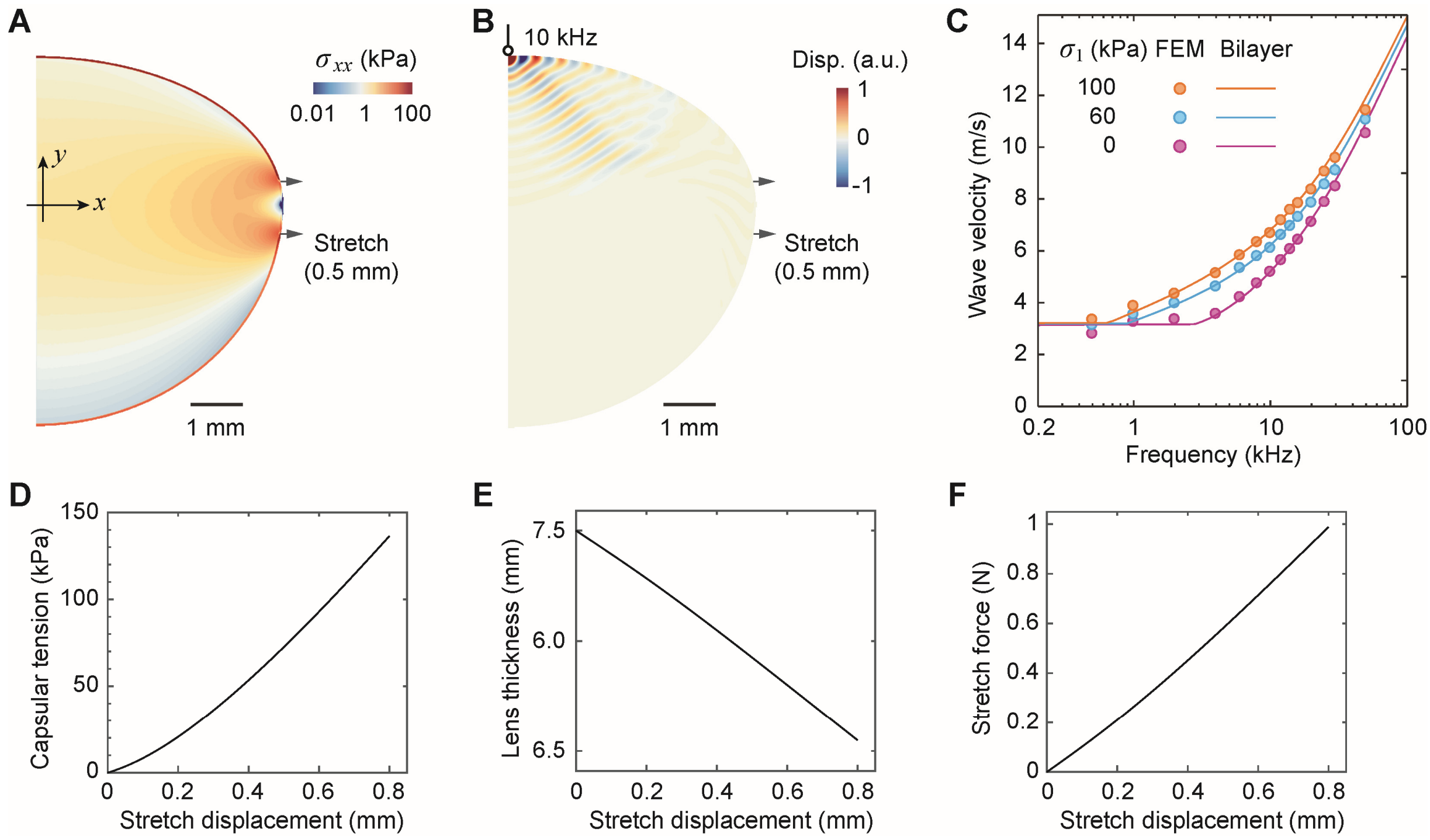

**Fig. 5**. Finite element lens model under zonular stretch. (**A**) Distribution of lateral stress ($\sigma_{xx}$) in the lens when the upper and lower equatorial points are stretched by a displacement of 0.5 mm for $\mu_1$ = 500 kPa and $\mu_2$ = 10 kPa. (**B**) Wave-field generated at an excitation frequency of 10 kHz. (**C**) Dispersion curves under different capsular tensions: 0, 60 kPa and 100 kPa, induced by a zonular stretch of 0, 450 μm, and 650 μm, respectively. Circles represent FEA results, while solid curves denote bilayer model results. Purple curve: $\sigma_1 = \sigma_2 = 0$, $h$ = 50 μm. Blue curve: $\sigma_1$ = 60 kPa, $\sigma_2$ = 0.6 kPa, $h$ = 48 μm. Orange curve: $\sigma_1$ = 100 kPa, $\sigma_2$ = 1 kPa, $h$ = 46 μm. (**D**) Relationship between anterior capsular tension and applied stretch. (**E**) Relationship between lens thickness and applied stretch. (**F**) Relationship between force and applied stretch.

### 3.5. Quantification of capsular tension induced by biaxial stretching

To simulate the accommodation process, we stretched the intact lens within the eye globe using a biaxial stretcher (Fig. 6A). Figure 6B provides a schematic representation of the lens in both non-stretch (relaxed) and stretched conditions. The non-stretch condition corresponds to maximum accommodation, where zonular fiber forces are released, while the stretched condition simulates the lens under tension from zonule fiber forces [4, 33]. OCT images (Fig. 6C) reveal significant flattening of the anterior lens surface under axial stretch, estimated at approximately 4%. This strain value represents the physiological range of the human lens under accommodation [52]. Upon stretching, the wavelength of elastic waves increased markedly (Fig. 6D). Figure 6E shows the measured wave velocities in both relaxed and stretched conditions. We performed simultaneous curve fitting of the two dispersion datasets, assuming identical shear modulus but variable tension across conditions. The fitting yielded a capsular shear modulus ($\mu_1$) of 570 ± 38 kPa for the anterior capsule. Capsular tension ($\sigma_1$) increased from 8 ± 6 kPa in the relaxed state and to 75 ± 4 kPa in the stretched condition, while cortical tension ($\sigma_2$) rose from 0.1 ± 0.09 Pa to 1.1 ± 0.07 Pa. The resulting stretch-induced changes in tension were $\Delta\sigma_1$ = 67 ± 2 kPa and $\Delta\sigma_2$ = 1 ± 0.05 kPa. Notably, the relatively low uncertainty in $\Delta\sigma_1$ (± 2 kPa), despite larger uncertainties in individual values, arises from the correlated nature of the tension parameters across the two conditions.

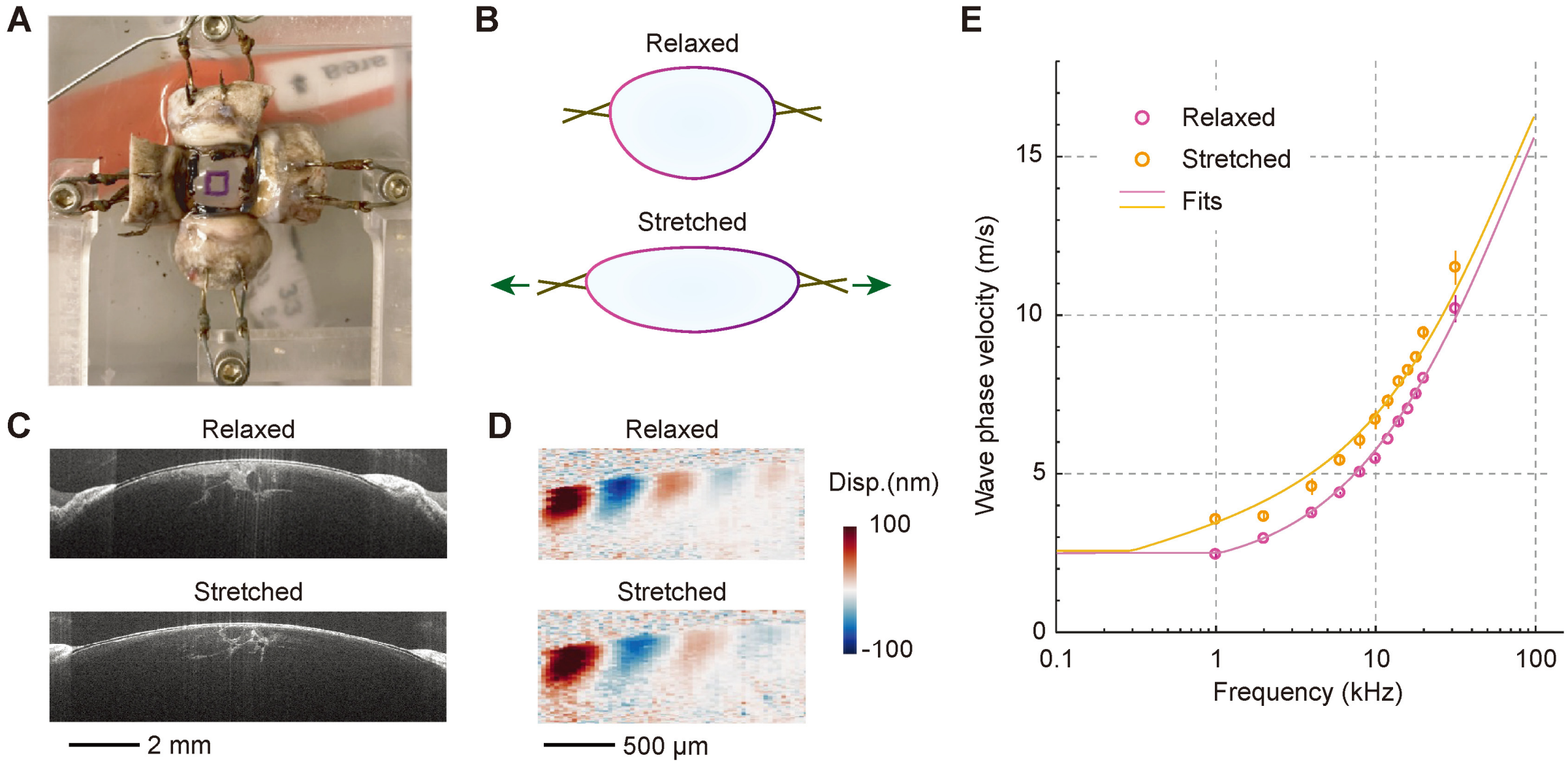

**Fig. 6**. Biaxial stretching of an intact porcine lens. (**A**) Photograph of an eye globe mounted in a biaxial stretching device, applying tensional force to the surrounding sclera and ciliary zonular fibers. (**B**) Schematic representation of the lens in relaxed and stretched conditions. (**C**) OCT images showing the anterior lens in relaxed and stretch conditions. (**D**) OCE wave profiles at 6 kHz in relaxed and stretched conditions. (**E**) Wave velocities of the anterior lens in relaxed and stretched conditions. Circles represent the experimental mean values from three measurements at different locations on the sample, with error bars denoting standard deviation. The best-fit parameters were $h$ = 55 µm, $\mu_1$ = 570 kPa, and $\mu_2$ = 6 kPa for both conditions. Tensions were $\sigma_1$ = 8 kPa and $\sigma_2$ = 0.15 kPa for the relaxed state, and $\sigma_1$ = 75 kPa and $\sigma_2$ = 1.1 kPa for the stretched state.

## 4. Discussion

Our results demonstrate the effectiveness of deriving mechanical parameters from leaky surface waves, which are supported in layered structures with a stiff surface and softer interior. These highly dispersive waves exist between two critical frequencies, $f_{c2}$ and $f_{c1}$, and their dispersion curves exhibit rich features from which elastic moduli and internal tensions can be extracted. Leveraging a recently developed high-frequency, wideband OCE system, we successfully applied this approach to the lens capsule-tissue system, enabling quantification of both intrinsic and externally induced capsular tensions with kPa-scale resolution. To the best of our knowledge, this represents the first non-destructive measurement of lens capsular tension.

For typical lens parameters ($\mu_1$ = 500 kPa, $\mu_2$ = 3 kPa, $h$ = 55 μm, $\rho_i$ = 1000 kg/m$^3$ and $\sigma_1$ ranging from 0 to 60 kPa), the lower and upper cutoff frequencies $f_{c2}$ and $f_{c1}$ are estimated to be 600-800 Hz and 400 kHz, respectively. Conventional OCE systems, limited to frequencies of 2-4 kHz, are inadequate for characterizing lens capsules. In this study, we optimized mechanical stimulation and system noise control to reliably capture wave velocities up to 30 kHz—critical for measuring capsular tension. The upper frequency limit in our system is currently constrained by signal-to-noise ratio (SNR), as wave displacement amplitude decreases and attenuation increases with frequency. Extending this limit could improve both the accuracy and precision of parameter estimation. While our prior work has shown that various stimulus waveforms (chirp, impulse, and monotone) can generate dispersion curves [53], the single-frequency excitation used here is well suited for visualizing displacement profiles and measuring wavelengths at specific frequencies.

Normal ILP in humans is thought to approximate intraocular pressure (IOP), ranging from 10 to 20 mmHg (1.3 to 2.6 kPa). A detailed study in murine lenses [54] using intracellular microelectrodes revealed a parabolic ILP profile, decreasing from ~330 mmHg (44 kPa) at the center to near zero at the surface. This pressure gradient drives intracellular fluid circulation through gap junction channels [54]. In our measurements on isolated six-month-old porcine lenses, free from external tension, we observed a posterior capsular tension ($\sigma_1$) of 44 kPa and a cortical tension ($\sigma_2$) of 800 Pa. Applying the Young-Laplace relation, $\sigma_1 = PR/2h$, with $h$ = 15 μm and $R$ = 6 mm, we estimate the hydrostatic ILP at the posterior surface to be ~220 Pa. Similarly, using $\sigma_2 = R/2\,(dP/dr)$, a posterior pressure gradient of ~260 Pa/mm is estimated. In contrast, the anterior capsule exhibited a lower tension of ~8 kPa. Substituting into the Young-Laplace relation with $h$ = 55 μm and $R$ = 6 mm, we estimate the anterior ILP to be ~150 Pa, comparable to the posterior value.

The shear moduli of the anterior and posterior porcine capsules were measured to be 630 ± 273 kPa and 440 ± 107 kPa, respectively, corresponding to Young's moduli of 1.89 ± 0.8 MPa and 1.32 ± 0.3 MPa. The higher modulus in the anterior capsule aligns with prior reports [5, 6], attributed to regional differences in macromolecular composition, including proteoglycans and fibronectin. Furthermore, the age-related loss of lamellar structure in the posterior capsule has been documented [5]. Our results fall within previously reported ranges for capsular Young's modulus: 1.5 – 6 MPa from spinning tests [16], and 0.3 – 6 MPa from tensile tests [6, 35, 55, 56].

Capsulorhexis, a common procedure in cataract surgery [5] involving removal of the anterior lens capsule, may benefit from preoperative assessment of capsule stiffness and tension to optimize surgical outcomes. Additionally, monitoring capsular tension following the placement of intraocular lens (IOL) implants could inform long-term prognosis and implant stability.

In our study, the lens was extracted from the intact globe and exposed to air. In vivo, the lens interfaces with aqueous and vitreous humors, resulting in different boundary conditions that influence wave dispersion. To address this, we extended the bilayer model to account for fluid contact (see Supplementary Note 4). Numerical simulations show that, for identical material properties, phase velocities in the fluid-contact condition are approximately 20% lower than those in the air-contact case across the 1–30 kHz range. As a result, using the air-contact model for in vivo data would underestimate both modulus and tension. Therefore, in vivo characterization of lens properties requires the fluid-contact bilayer model in conjunction with the broadband OCE method.

Although we employed an isotropic material model, lens tissue exhibits transverse isotropy, similar to corneal stroma [16, 57]. Incompressible, transversely isotropic materials can be described using three independent elastic constants: out-of-plane shear modulus $\mu_{xy}$ (= $\mu_{zy}$), in-plane shear modulus $\mu_{xz}$, and polar-direction Young's modulus $E_{yy}$. We derived a dispersion equation for a bilayer model with transversely isotropy (see Supplementary Note 5) and found that dispersion is primarily influenced by $\mu_{xy}$ of both layers, with minor contributions from $\mu_{xz}$ and $E_{yy}$. Hence, the shear moduli $\mu_1$ and $\mu_2$ used in our isotropic model effectively reflect $\mu_{xy}$ values. Given the moderate anisotropy in lens tissue ($\mu_{xy}/\mu_{xz}$ < 2) [16], the isotropic bilayer model introduces an error of less than 1% (Supplementary Note S5.2).

While our current model assumes purely elastic behavior, previous studies have demonstrated the viscoelastic nature of the lens and its influence on wave propagation [30, 58]. Future work should incorporate a viscoelastic bilayer model to more accurately characterize lens dynamics [59].

In our biaxial lens stretching experiment, the OCE system effectively detected an increase in anterior capsular tension of approximately 67 kPa under 4% strain, compared to the fully relaxed state, with a low uncertainty of only 2 kPa (3%). This corresponds to a strain sensitivity of 0.12%. Based on our FEM simulation (Fig. 5), this level of uncertainty in tension difference translates to sensitivities of 12 μm in lens stretch amplitude and of 14 mN in zonular stretch force. The current stretcher applies force to the sclera rather than directly engaging the zonule fibers. Refining the device to better mimic physiological accommodation could further enhance relevance to the human lens system [51]. Nevertheless, our results highlight the translational potential of OCE for clinical applications, including diagnosing and monitoring accommodative dysfunctions such as presbyopia and optimizing capsular surgery strategies and procedures.

Currently, our OCE system relies on contact-based mechanical stimulation. However, previous studies have shown that acoustic radiation force generated by focused ultrasound can induce elastic waves with sufficient displacement amplitude in the lens [29, 30]. Integrating non-contact ultrasound stimulation with wideband OCE could enable real-time, in vivo evaluation of the human lens, facilitating clinical translation.

Beyond ophthalmology, this optical elastography technique could prove valuable for measuring stress, strain, and mechanical properties in other biological tissues and biomaterials. Potential applications include assessing mechanical tension in blood vessels, tendons, ligaments, and skin.

**Funding.** National Institutes of Health (R01-EY033356, R01-EY034857).

**Acknowledgements.**

**Disclosures.** The authors declare no conflicts of interest.

**Data availability.** Data underlying the results presented in this paper are not publicly available at this time but may be obtained from the authors upon reasonable request.

**Code availability.** The code for computing wave dispersion has been made publicly available at: https://github.com/JanYxuan/Bilayer-wave-dispersion.git

**Appendix**

The components of the matrix $\mathbf{M}_{6\times 6}$ are as follows:

$$
\begin{aligned}
&M_{11} = s_1,\ M_{12} = s_2,\ M_{13} = -s_1,\ M_{14} = -s_2,\ M_{15} = -s_1^*,\ M_{16} = -s_2^*,\\
&M_{21} = 1,\ M_{22} = 1,\ M_{23} = 1,\ M_{24} = 1,\ M_{25} = -1,\ M_{26} = -1,\\
&M_{31} = \gamma_1(1+{s_1}^2),\ M_{32} = \gamma_1(1+{s_2}^2),\ M_{33} = \gamma_1(1+{s_1}^2),\ M_{34} = \gamma_1(1+{s_2}^2),\\
&M_{35} = -\gamma_2\left(1+s_1^{*2}\right),\ M_{36} = -\gamma_2\left(1+s_2^{*2}\right),
\end{aligned}
\tag{A.1}
$$

$M_{41} = \gamma_1 s_1(1 + {s_2}^2)$, $M_{42} = \gamma_1 s_2(1 + {s_1}^2)$, $M_{43} = -\gamma_1 s_1(1 + {s_2}^2)$,

$M_{44} = -\gamma_1 s_2(1 + {s_1}^2)$, $M_{45} = -\gamma_2 s_1^*(1 + s_2^{*2})$, $M_{46} = -\gamma_2 s_2^*(1 + s_1^{*2})$,

$M_{51} = (1 + {s_1}^2)\exp(s_1 kh)$, $M_{52} = (1 + {s_2}^2)\exp(s_2 kh)$, $M_{53} = (1 + {s_1}^2)\exp(-s_1 kh)$,

$M_{54} = (1 + {s_2}^2)\exp(-s_2 kh)$, $M_{55} = 0$, $M_{56} = 0$,

$M_{61} = s_1(1 + {s_2}^2)\exp(s_1 kh)$, $M_{62} = s_2(1 + {s_1}^2)\exp(s_2 kh)$,

$M_{63} = -s_1(1 + {s_2}^2)\exp(-s_1 kh)$, $M_{64} = -s_2(1 + {s_1}^2)\exp(-s_2 kh)$, $M_{65} = 0$, $M_{66} = 0$.

Here, $\gamma_1 = \mu_1 - 2\sigma_1/3$, $\gamma_2 = \mu_2 - 2\sigma_2/3$; $s_1$ and $s_2$ are the roots of the wave equation in the first layer:

$$\left(\mu_1 - \frac{2}{3}\sigma_1\right) s^4 - \left(2\mu_1 - \frac{1}{3}\sigma_1 - \rho_1 \frac{\omega^2}{k^2}\right) s^2 + \mu_1 + \frac{1}{3}\sigma_1 - \rho_1 \frac{\omega^2}{k^2} = 0; \tag{A.2}$$

and $s_1^*$ and $s_2^*$ are the roots of the wave equation in the second layer:

$$\left(\mu_2 - \frac{2}{3}\sigma_2\right) s^{*4} - \left(2\mu_2 - \frac{1}{3}\sigma_2 - \rho_2 \frac{\omega^2}{k^2}\right) s^{*2} + \mu_2 + \frac{1}{3}\sigma_2 - \rho_2 \frac{\omega^2}{k^2} = 0, \tag{A.3}$$

where $\rho_1$ and $\rho_2$ denote the densities of the two layers, $\omega$ ($= 2\pi f$) is the angular frequency, and $k$ is the wavenumber.

**Supplementary materials.** Supplementary material associated with this article can be found in the online version.

**Supplementary Information**

# Optical Coherence Elastography Measures Mechanical Tension in the Lens and Capsule

Xu Feng[1,4,†], Guo-yang Li[1,3,†], Yuxuan Jiang[1,†], Owen Shortt-Nguyen[1], Seok-Hyun Yun[1,2,*]

[1] Harvard Medical School and Wellman Center for Photomedicine, Massachusetts General Hospital, 50 Blossom St., Boston, MA 02114, USA.

[2] Harvard-MIT Division of Health Sciences and Technology, Cambridge, MA 02139, USA.

[3] Current address: Department of Mechanics and Engineering Science, College of Engineering, Peking University, Beijing 100871, China.

[4] Current address: Department of Bioengineering, University of Texas at Dallas, TX 75080, USA.

[†] Equal contribution

*Correspondence: syun@hms.harvard.edu

**Supplementary Note 1. Derivation of the tension-induced bilayer guided wave model**

Consider a solid material that is subjected to finite deformation, and infinitesimal elastic waves are superimposed on the static deformation. The waves can be described by the following linear equation of motion [1, 2]:

$$\nabla \cdot \boldsymbol{\Sigma} = \rho \boldsymbol{u}_{,tt}, \tag{S1}$$

where $\boldsymbol{\Sigma}$ denotes incremental stress induced by elastic waves, $\boldsymbol{u}$ denotes the displacement of wave motion, $\rho$ denotes the mass density of material, $t$ denotes the time, and subscript comma represents derivative with respect to spatial coordinates. The harmonic elastic wave can be described as $\boldsymbol{u} = \boldsymbol{u}_0 \exp(i(\boldsymbol{k} \cdot \boldsymbol{x} - \omega t))$. where $\boldsymbol{u}_0$, $\boldsymbol{k}$ and $\omega$ denote wave amplitude, wave vector, and angular frequency, respectively. For incompressible materials, the incremental stress $\boldsymbol{\Sigma}$ is related to the displacement by $\Sigma_{ij} = \mathcal{A}^0_{ijkl} u_{l,k} - \hat{p}\delta_{ij} + p u_{i,j}$, where $\hat{p}$ denotes the increment of the Lagrange multiplier $p$. $\mathcal{A}^0_{ijkl}$ is the fourth-order Eulerian elasticity tensor defined as [1, 2]:

$$\mathcal{A}^0_{ijkl} = F_{iI} F_{kJ} \frac{\partial^2 W}{\partial F_{jI} \partial F_{lJ}}, \qquad i, j, k, l, I, J \in \{x, y, z\} \tag{S2}$$

where $\boldsymbol{F} = \mathrm{diag}(\lambda_x, \lambda_y, \lambda_z)$ is the deformation gradient tensor, and $W$ is the strain energy function of the material. When the wave has displacement components only within the $x$-$y$ plane (i.e., $u_z = 0$), the stream function $\psi(x, y, t)$ can be used to replace displacements: $u_x = \psi_{,y}$ and $u_y = -\psi_{,x}$. Inserting $\psi$ and $\mathcal{A}^0_{ijkl}$ into Eq. (S1), the wave equation becomes

$$\alpha \psi_{,xxxx} + 2\beta \psi_{,xxyy} + \gamma \psi_{,yyyy} = \rho(\psi_{,xxtt} + \psi_{,yytt}), \tag{S3}$$

where $\alpha$, $\beta$, and $\gamma$ are mechanical parameters given by $\alpha = \mathcal{A}^0_{xyxy}$, $2\beta = \mathcal{A}^0_{xxxx} + \mathcal{A}^0_{yyyy} - 2\mathcal{A}^0_{xxyy} - 2\mathcal{A}^0_{xyyx}$, and $\gamma = \mathcal{A}^0_{yxyx}$.

A flat bilayer model consists of a top layer (layer 1) with a thickness $h$ and a semi-infinite bottom layer (layer 2). The $y$-axis denotes the thickness direction, and the $x$-axis aligns with the direction of wave propagation along the layer. The stream function of the top layer is described as: $\psi = \psi_0 \exp(sky) \exp[i(kx - \omega t)]$, where $\psi_0$ is an amplitude and $s$ denotes the ratio of wavenumbers between the $y$ and $x$ directions. Inserting $\psi$ into Eq. (S3) gives

$$\gamma_1 s^4 - \left(2\beta_1 - \rho_1 \frac{\omega^2}{k^2}\right) s^2 + \alpha_1 - \rho_1 \frac{\omega^2}{k^2} = 0. \quad \text{(S4)}$$

The stream function in the layer 2 is $\psi^*$ $(= \psi_0^* \exp(s^* k y) \exp[i(kx - \omega t)])$. Inserting it into Eq. (S3) gives

$$\gamma_2 s^{*4} - \left(2\beta_2 - \rho_2 \frac{\omega^2}{k^2}\right) s^{*2} + \alpha_2 - \rho_2 \frac{\omega^2}{k^2} = 0, \quad \text{(S5)}$$

where $s^*$ is the ratio of wavenumbers between the two directions. $\rho_1$ and $\rho_2$ denote the density of the top layer and the substrate, respectively. The incremental parameters of the top layer $\alpha_1$, $\beta_1$, and $\gamma_1$ are given by the strain energy function of the top layer, $W^{(1)}$, via Eq. (S2). The incremental parameters of the substrate $\alpha_2$, $\beta_2$, and $\gamma_2$ are given by the strain energy function of the substrate $W^{(2)}$. The interface of the two layers (at $y = 0$) ensures continuity of displacement and stress, and the top surface of the layer exposed to air (at $y = h$) is stress free. These boundary conditions can be written as:

$$u_x = u_x^*, u_y = u_y^*, \Sigma_{yx} = \Sigma_{yx}^*, \Sigma_{yy} = \Sigma_{yy}^*, \text{at } y = 0;$$
$$\Sigma_{yx} = 0, \Sigma_{yy} = 0, \text{ at } y = h, \quad \text{(S6)}$$

where $u_i$ and $u_i^*$ denote the displacement in the top layer and the substrate, respectively, and $\Sigma_{ij}$ and $\Sigma_{ij}^*$ denote the incremental stress in the top layer and the substrate, respectively. Taking $\psi$ and $\psi^*$ into the boundary conditions, a secular equation can be obtained as [3]:

$$\det(\mathbf{M}_{6\times6}) = 0, \quad \text{(S7)}$$

where $\mathbf{M}_{6\times6}$ is a 6×6 matrix with the following components:

$M_{11} = s_1,\ M_{12} = s_2,\ M_{13} = -s_1,\ M_{14} = -s_2,\ M_{15} = -s_1^*,\ M_{16} = -s_2^*,$

$M_{21} = 1,\ M_{22} = 1,\ M_{23} = 1,\ M_{24} = 1,\ M_{25} = -1,\ M_{26} = -1,$

$M_{31} = \gamma_1(1 + {s_1}^2),\ M_{32} = \gamma_1(1 + {s_2}^2),\ M_{33} = \gamma_1(1 + {s_1}^2),$

$M_{34} = \gamma_1(1 + {s_2}^2),\ M_{35} = -\gamma_2\left(1 + s_1^{*2}\right),\ M_{36} = -\gamma_2\left(1 + s_2^{*2}\right),$

$M_{41} = \gamma_1 s_1(1 + {s_2}^2),\ M_{42} = \gamma_1 s_2(1 + {s_1}^2),\ M_{43} = -\gamma_1 s_1(1 + {s_2}^2),$

$M_{44} = -\gamma_1 s_2(1 + {s_1}^2),\ M_{45} = -\gamma_2 s_1^*\left(1 + s_2^{*2}\right),\ M_{46} = -\gamma_2 s_2^*\left(1 + s_1^{*2}\right),$

$$M_{51} = (1 + {s_1}^2)\exp(s_1 kh),\ M_{52} = (1 + {s_2}^2)\exp(s_2 kh),$$

$$M_{53} = (1 + {s_1}^2)\exp(-s_1 kh),\ M_{54} = (1 + {s_2}^2)\exp(-s_2 kh),\ M_{55} = 0,\ M_{56} = 0,$$

$$M_{61} = s_1(1 + {s_2}^2)\exp(s_1 kh),\ M_{62} = s_2(1 + {s_1}^2)\exp(s_2 kh),$$

$$M_{63} = -s_1(1 + {s_2}^2)\exp(-s_1 kh),\ M_{64} = -s_2(1 + {s_1}^2)\exp(-s_2 kh),$$

$$M_{65} = 0,\ M_{66} = 0. \quad \text{(S8)}$$

When the bilayer system is subjected to equal and small strain along the *x* and *z* axes (i.e. $\varepsilon = \varepsilon_x = \varepsilon_z$), the stretch ratios are simplified to $\lambda_x = \lambda_z = 1 + \varepsilon,\ \lambda_y = 1 - 2\varepsilon$. With the small strain assumption ($\varepsilon \ll 1$), the linear approximation of the incremental parameters can be obtained as follows: (this can be easily verified by employing commonly used strain energy functions for tissues, such as the Neo-Hookean or Fung model [4], along with the small-strain relation substituted into $\mathcal{A}^0_{ijkl}$ and $\mathcal{A}^{0*}_{ijkl}$.)

$$\begin{aligned}
\alpha_1 &= \mu_1(1 + 2\varepsilon) = \mu_1 + \tfrac{1}{3}\sigma_1,\\
\gamma_1 &= \mu_1(1 - 4\varepsilon) = \mu_1 - \tfrac{2}{3}\sigma_1,\\
\beta_1 &= \mu_1 - \tfrac{1}{6}\sigma_1,
\end{aligned} \quad \text{(S9)}$$

and

$$\begin{aligned}
\alpha_2 &= \mu_2(1 + 2\varepsilon) = \mu_2 + \tfrac{1}{3}\sigma_2,\\
\gamma_2 &= \mu_2(1 - 4\varepsilon) = \mu_2 - \tfrac{2}{3}\sigma_2,\\
\beta_2 &= \mu_2 - \tfrac{1}{6}\sigma_2,
\end{aligned} \quad \text{(S10)}$$

where $\sigma_1\ (= \sigma_{1x} = \sigma_{1z})$ and $\sigma_2\ (= \sigma_{2x} = \sigma_{2z})$ represent biaxial stresses in the top and bottom layers, and $\mu_1$ and $\mu_2$ denote the shear moduli of the top and bottom layers. Inserting Eqs. (S9) and (S10) into Eq. (S7), the dispersion equations (A.1) - (A.3) in Appendix are obtained.

**Supplementary Note 2. The effect of intra-lenticular pressure gradient on cortical stress**

The intra-lenticular pressure within the lens is assumed to depend only on the radial coordinate: $P = P(r)$. The radial equilibrium equation is given by [5]:

$$\frac{1}{r^2}\frac{\partial(r^2\sigma_{rr})}{\partial r}+\frac{1}{r\sin\theta}\frac{\partial(\sigma_{r\theta}\sin\theta)}{\partial\theta}+\frac{1}{r\sin\theta}\frac{\partial\sigma_{r\phi}}{\partial\phi}-\frac{\sigma_{\theta\theta}+\sigma_{\phi\phi}}{r}=0, \tag{S11}$$

where radial stress $\sigma_{rr} = -P(r)$. Inserting $\sigma_{rr}$ into Eq. (S11) gives

$$\begin{aligned}\sigma_{r\theta} &= \sigma_{r\phi} = 0,\\ \sigma_{\theta\theta} &= \sigma_{\phi\phi} = -P(r) - \frac{r}{2}\frac{\mathrm{d}P(r)}{\mathrm{d}r}.\end{aligned} \tag{S12}$$

A variation of the hydrostatic pressure along the radial coordinate gives rise to deviatoric tensile stress $\sigma_2$ within the cortex. It can be shown that $\sigma_2(r) = \frac{r}{2}\frac{\mathrm{d}P}{\mathrm{d}r}$.

**Supplementary Note 3. Critical frequencies between nonleaky and leaky branches**

For a homogeneous material, the Rayleigh surface wave of a material in the pre-stressed state satisfies the following relation [6]:

$$\sqrt{(\alpha - \rho {v_R}^2)/\gamma} \approx 0.2956, \tag{S13}$$

Inserting Eqs. (S9) and (S10) into Eq. (S13) leads to

$$\rho_i {v_{Ri}}^2 \approx 0.913\, \mu_i + 0.392\, \sigma_i \quad (i = 1, 2). \tag{S14}$$

The first term in Eq. (S14) is the well-known Rayleigh wave velocity in the stress-free condition. The second term is the effect of tension.

For a bilayer material consisting of a stiff layer and a soft substrate, there exists a critical frequency $f_c$ that separates wave dispersion into nonleaky and leaky branches [7, 8]. The phase velocity at the critical frequency is equal to the bulk shear wave velocity in the substrate, $v_{T2}$, where $v_{T2} = \sqrt{\alpha_2/\rho_2} = \sqrt{(\mu_2 + \sigma_2/3)/\rho_2}$.

There are nine variables involved in this physical system, namely $\mu_1$, $\mu_2$, $\sigma_1$, $\sigma_2$, $\rho_1$, $\rho_2$, $h_1(= h)$, and $h_2(\to \infty)$, as well as the cutoff frequency $f_c$. These variables depend on the three physical dimensions: mass, length and time. Therefore, according to the Buckingham pi theorem [9], six dimensionless parameters characterize the system. We may consider these six parameters comprising of a dimensionless critical frequency $hf_c/v_{T2}$, dimensionless modulus $\mu_2/\mu_1$, dimensionless stresses $\sigma_1/\mu_1$ and $\sigma_2/\mu_1$, dimensionless density $\rho_2/\rho_1$, and dimensionless wall thickness $h_1/h_2$ ($\approx 0$). The dimensionless critical frequency can then be expressed in terms of four dimensionless parameters as follows:

$$\frac{hf_c}{v_{T2}} = \Pi(\frac{\mu_2}{\mu_1}, \frac{\sigma_1}{\mu_1}, \frac{\sigma_2}{\mu_1}, \frac{\rho_2}{\rho_1}), \tag{S15}$$

where $\Pi$ denotes an implicit dimensionless function. $\rho_2/\rho_1$ is a constant close to 1. Given that $\sigma_2 \ll \mu_2 \ll \mu_1$, the effect of $\sigma_2/\mu_1$ on the dimensionless frequency is negligible. Eq. (S15) can be further simplified to

$$\frac{hf_c}{v_{T2}} = \Pi(\frac{\mu_2}{\mu_1}, \frac{\sigma_1}{\mu_1}), \tag{S16}$$

We numerically solved the secular equation (S6) and obtained $\frac{hf_c}{v_{T2}}$ as a function of $\frac{\mu_2}{\mu_1}$ and $\frac{\sigma_1}{\mu_1}$. Figure S1a shows the sensitivity of the dimensionless frequency to the two dimensionless parameters.

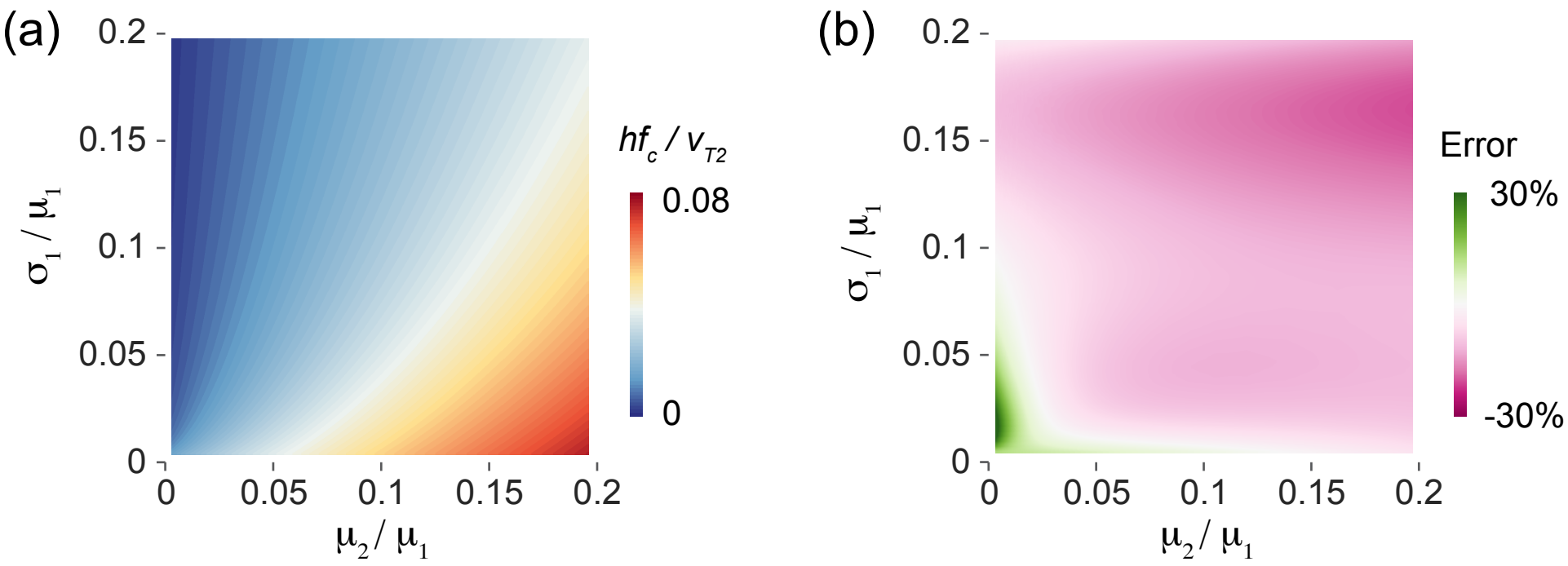

**Fig. S1**. (**a**) Variation of the dimensionless critical frequency $hf_c/v_{T2}$ with respect to dimensionless parameters $\mu_2/\mu_1$ and $\sigma_1/\mu_1$. (**b**) The relative error between the theoretical solution and the fitting function. In most regions, the relative error is less than 15%.

By fitting the dimensionless function, the following approximation expressions are found:

$$\frac{hf_c}{v_{T2}} = C_1 \left(\frac{\mu_2}{\mu_1}\right)^{C_2}, \tag{S17}$$

where

$$C_1 = 73.7\left(\frac{\sigma_1}{\mu_1}\right)^3 - 26.8\left(\frac{\sigma_1}{\mu_1}\right)^2 + 2.4\frac{\sigma_1}{\mu_1} + 0.16,$$
$$C_2 = 1 - 0.5\exp\left(-30\frac{\sigma_1}{\mu_1}\right). \tag{S18}$$

It can be expressed in terms of Young's moduli, $E_1 = 3\mu_1$ and $E_2 = 3\mu_2$:

$$\frac{hf_c}{v_{T2}} = C_1 \left(\frac{E_2}{E_1}\right)^{C_2}, \tag{S19}$$

where

$$C_1 = 1989.9\left(\frac{\sigma_1}{E_1}\right)^3 - 241.2\left(\frac{\sigma_1}{E_1}\right)^2 + 7.2\frac{\sigma_1}{E_1} + 0.16,$$
$$C_2 = 1 - 0.5\exp\left(-90\frac{\sigma_1}{E_1}\right). \tag{S20}$$

In the stress-free condition, Eq. (S19) reduces to $hf_c/v_{T2} = 0.16\sqrt{E_2/E_1}$ [8]. Figure S1b shows the relative error between the dimensionless critical frequencies obtained from the theoretical solution and the fitting function. Eqs. (S17) and (S17) provide a good approximation when $\mu_2/\mu_1 \leq 0.2$ and $\sigma_1/\mu_1 \leq 0.2$.

### Supplementary Note 4. Bilayer model including fluid

### S4.1 Secular equation of the bilayer model immersed in fluid

Figure S2a shows the schematic of a bilayer model immersed in fluid. The surface of Layer 1 is in contact with fluid. The interface between Layer 1 and fluid (at $y = h$) satisfies the continuity of normal displacement and normal stress and the condition of zero shear stress. The interface between Layer 1 and Layer 2 (at $y = 0$) ensures continuity of displacement and stress. These boundary conditions can be written as:

$$\begin{gathered} u_x = u_x^*, u_y = u_y^*, \Sigma_{yx} = \Sigma_{yx}^*, \Sigma_{yy} = \Sigma_{yy}^*, \text{at } y = 0; \\ u_y = u_y^f, \Sigma_{yx} = -\sigma_{yy} u_{y,x}, \Sigma_{yy} = -p^f - \sigma_{yy} u_{y,y}, \text{ at } y = h, \end{gathered} \tag{S21}$$

where $\boldsymbol{u}^f$ denotes the displacement of the fluid, and $p^f$ is the fluid hydrostatic pressure.

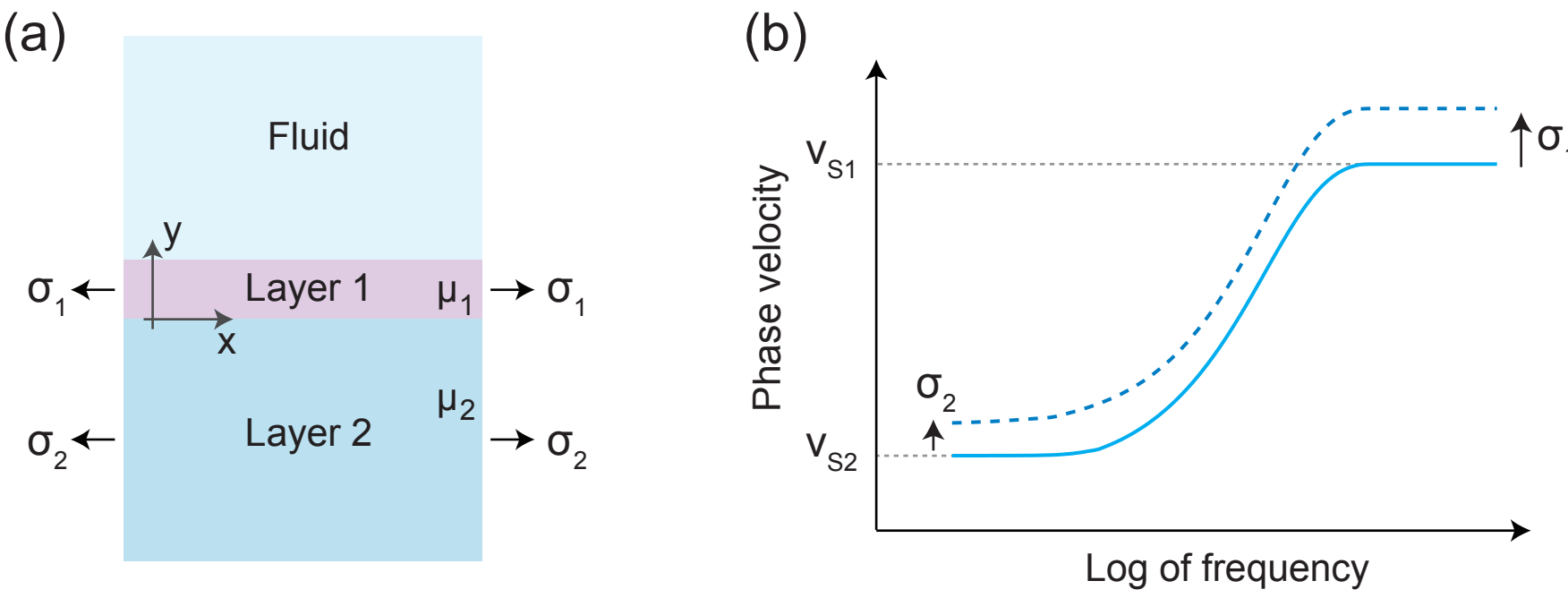

**Fig. S2**. Bilayer model in contact with fluid. (**a**) Schematic of the bilayer model. (**b**) Representative dispersion curves in the stress-free and stressed states, respectively.

Using the same methods involving stream functions and harmonic waveforms as described in Supplementary Note 1, we obtain a secular equation [10]:

$$\det(\mathbf{M}_{7\times 7}) = 0, \tag{S22}$$

where $\mathbf{M}_{7\times 7}$ is a 7×7 matrix with the following components:

$$M_{11} = s_1,\ M_{12} = s_2,\ M_{13} = -s_1,\ M_{14} = -s_2,\ M_{15} = -s_1^*,\ M_{16} = -s_2^*,\ M_{17} = 0,$$

$$M_{21} = 1,\ M_{22} = 1,\ M_{23} = 1,\ M_{24} = 1,\ M_{25} = -1,\ M_{26} = -1,\ M_{27} = 0,$$

$$M_{31} = \gamma_1(1 + {s_1}^2),\ M_{32} = \gamma_1(1 + {s_2}^2),\ M_{33} = \gamma_1(1 + {s_1}^2),$$

$$M_{34} = \gamma_1(1+{s_2}^2),\ M_{35} = -\gamma_2\left(1+{s_1^*}^2\right),\ M_{36} = -\gamma_2\left(1+{s_2^*}^2\right),\ M_{37} = 0,$$

$$M_{41} = \gamma_1 s_1(1+{s_2}^2),\ M_{42} = \gamma_1 s_2(1+{s_1}^2),\ M_{43} = -\gamma_1 s_1(1+{s_2}^2),$$

$$M_{44} = -\gamma_1 s_2(1+{s_1}^2),\ M_{45} = -\gamma_2 s_1^*\left(1+{s_2^*}^2\right),\ M_{46} = -\gamma_2 s_2^*\left(1+{s_1^*}^2\right),\ M_{47} = 0,$$

$$M_{51} = \exp(s_1 kh),\ M_{52} = \exp(s_2 kh),\ M_{53} = \exp(-s_1 kh),$$

$$M_{54} = \exp(-s_2 kh),\ M_{55} = M_{56} = 0,\ M_{57} = i\xi\exp(-\xi kh),$$

$$M_{61} = (1+{s_1}^2)\exp(s_1 kh),\ M_{62} = (1+{s_2}^2)\exp(s_2 kh),$$

$$M_{63} = (1+{s_1}^2)\exp(-s_1 kh),\ M_{64} = (1+{s_2}^2)\exp(-s_2 kh),$$

$$M_{65} = M_{66} = M_{67} = 0,$$

$$M_{71} = s_1(1+{s_2}^2)\exp(s_1 kh),\ M_{72} = s_2(1+{s_1}^2)\exp(s_2 kh),$$

$$M_{73} = -s_1(1+{s_2}^2)\exp(-s_1 kh),\ M_{74} = -s_2(1+{s_1}^2)\exp(-s_2 kh),\ M_{75} = M_{76} = 0,$$

$$M_{77} = i\rho_f(\omega^2/k^2)\exp(-\xi kh). \qquad \text{(S23)}$$

Here, $s_1$ and $s_2$ are the two roots of Eq. (S4), and $s_1^*$ and $s_2^*$ are the two roots of Eq. (S5). The symbol $i$ in $M_{57}$ and $M_{77}$ represents the imaginary unit. $\xi = \sqrt{1 - \frac{1}{{c_f}^2}\frac{\omega^2}{k^2}}$, where $c_f$ is the acoustic speed in the fluid, given by ${c_f}^2 = \kappa_f/\rho_f$, where $\kappa_f$ is the bulk modulus of fluid (~2.2 GPa) and $\rho_f$ is the fluid density (~1000 kg/m$^3$).

Figure S2b show theoretical dispersion curves for two different stress values. The phase velocity at low frequencies correspond to the Scholte wave velocity of Layer 2 ($v_{S2}$). The phase velocity increases with frequency and eventually reaches a high-frequency plateau that corresponds to the Scholte wave velocity of Layer 1 ($v_{S1}$). An increase of stress $\sigma_2$ in Layer 2 raises the low-frequency plateau, while an increase of stress $\sigma_1$ in Layer 1 elevates the high-frequency plateau.

It can be shown that the Scholte wave velocities are approximately given by

$$\rho_i {v_{Si}}^2 \approx 0.704\,\mu_i + 0.530\,\sigma_i \quad (i = 1, 2). \qquad \text{(S24)}$$

Comparing these Scholte-wave asymptotes to the Rayleigh-wave asymptotes in Eq. (S14) for the air-interfacing model, the Scholte wave has a lower velocity but higher sensitivity to tension.

### S4.2 Comparison of the bilayer models exposed to air and fluid

Figure S3 compares the dispersion curves of the bilayer models exposed to air and fluid. The shear moduli of the layers 1 and 2 remain consistent in the two cases. The phase velocity of the air-contact model is always higher than that of the fluid-contact model. The differences in wave velocities across the frequency range of 1 – 30 kHz is ~ 20%. For lenses surrounded by the aqueous and vitreous humors within eyeballs, the fluid-contact bilayer model should be used instead of the air-contact model described in Supplementary Note S1. For the fluid-contact lenses, the air-contact model can underestimate shear modulus and stress by approximately 40%.

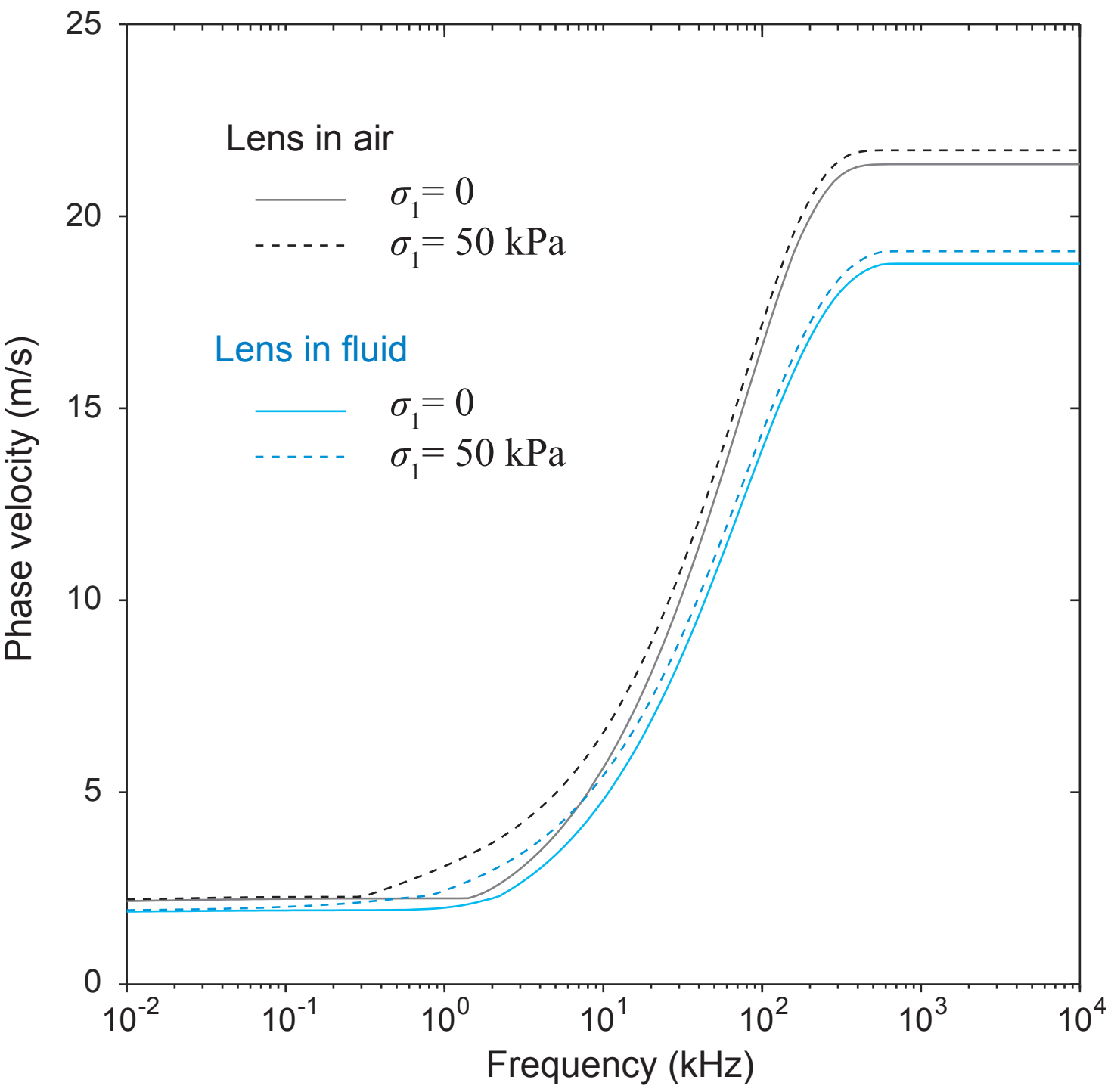

**Fig. S3**. Dispersion curves of a bilayer interfacing with air (black curves) or fluid (blue curves). The parameters used in this calculation are: $\mu_1 = 500$ kPa, $\mu_2 = 5$ kPa, $h = 55$ μm, $\rho_1 = \rho_2 = 1000$ kg/m$^3$, $\rho_f = 1000$ kg/m$^3$, $\kappa_f = 2.2$ GPa, and $\sigma_2 = 0.5$ kPa. Two different stress conditions are shown: $\sigma_1 = 0$ kPa (solid curves) and $\sigma_1 = 50$ kPa (dashed curves).

## Supplementary Note 5. Bilayer model for transversely isotropic materials

### S5.1 Derivation of the bilayer model solutions with transversely isotropy

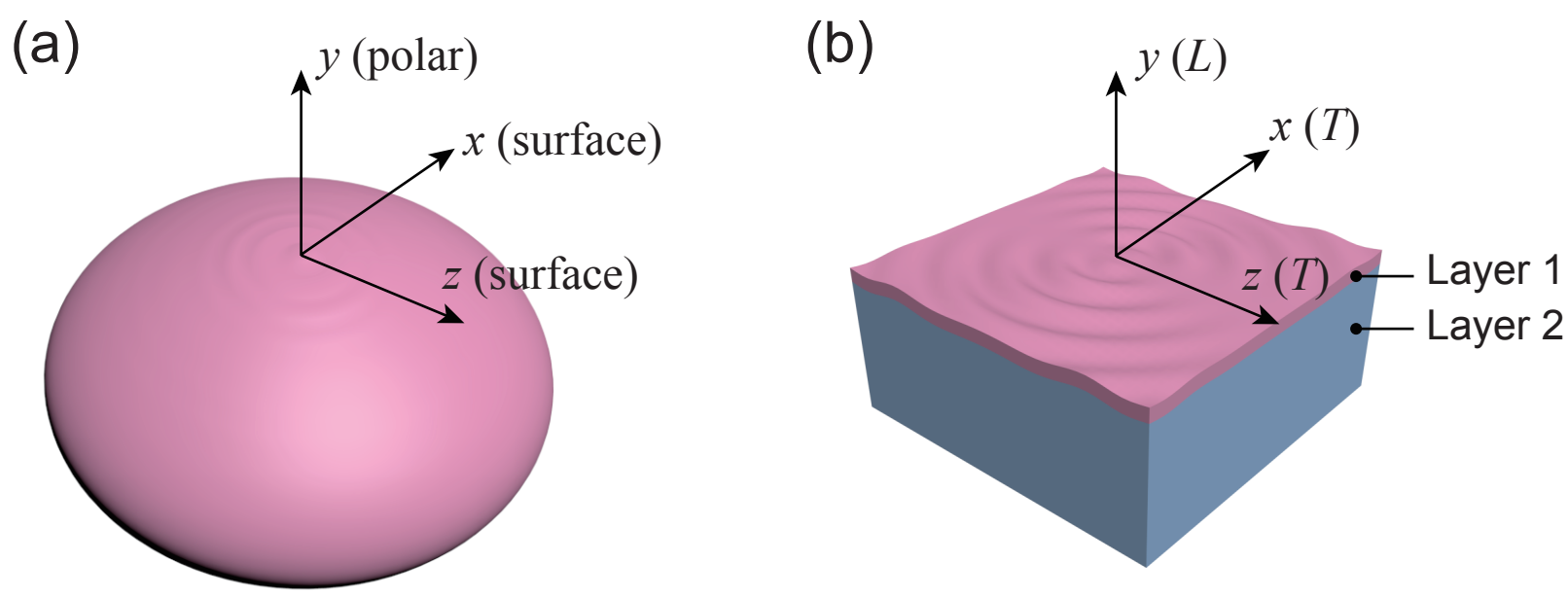

**Fig. S4**. Schematic of the lens model with transversely isotropic materials. (**a**) Ellipsoid-shape lens model. A Cartesian coordinate system ($x$, $y$, $z$) was built on the model. The $x$-$z$ plane is tangential to the capsule surface. The $y$-axis corresponds to the polar direction of the lens. (**b**) A flat bilayer model.

Figure S4a shows a schematic of a lens, where the $x$-$z$ plane is tangential to the lens surface, while the $y$-axis is normal to the surface plane (polar direction). In a flattened model (Fig. S4b) [11], the $x$-$z$ plane represents the transverse plane, and the y-axis corresponds to the longitudinal direction. The elastic waves propagate in the transverse plane. For a general transversely isotropic material, there are five independent elastic constants, including Young's modulus along the longitudinal direction $E_L$, Young's modulus in the transverse plane $E_T$, shear modulus along the longitudinal direction $\mu_L$, shear modulus in the transverse plane $\mu_T$, and Poisson's ratio $\nu_{LT}$. When the material is incompressible, there are two constraint relations:

$$\nu_{LT} = \frac{1}{2}, \; E_T = \frac{4\mu_T E_L}{\mu_T + E_L}, \tag{S25}$$

Therefore, the number of independent elastic constants is reduced to three: $\mu_L$, $\mu_T$, and $E_L$.

Several constitutive models have been used to describe transverse isotropic materials. Here let us consider the generalized Humphrey-Yin model [12, 13] with the strain energy function given by:

$$W = \frac{\mu_T}{2c_2}\left[e^{c_2(I_1-3)} - 1\right] + \frac{E_L+\mu_T-4\mu_L}{2c_4}\left[e^{c_4\left(\sqrt{I_4}-1\right)^2} - 1\right] + \frac{\mu_T-\mu_L}{2}(2I_4 - I_5 - 1), \tag{S26}$$

where invariants $I_4 = \lambda_y^2$, $I_5 = \lambda_y^4$. $c_2$ and $c_4$ are two dimensionless parameters denoting material's nonlinear hardening. Inserting $\lambda_x = \lambda_z = 1 + \varepsilon$, $\lambda_y = 1 - 2\varepsilon$, and $W$ into the definition of the incremental parameters, we obtain

$$\begin{aligned} \alpha &= \mu_L + (4\mu_T - 2\mu_L)\varepsilon, \\ \gamma &= \mu_L + (4\mu_T - 2\mu_L - 2E_L)\varepsilon, \\ \beta &= -\mu_L + \frac{E_L}{2} + \frac{\mu_T}{2} + (15\mu_T - 10\mu_L - 2E_L)\varepsilon. \end{aligned} \tag{S27}$$

Due to transverse isotropy, the elastic waves propagating along the *x*- and *z*-axes are governed by the same incremental parameters. The biaxial stress $\sigma = \sigma_x = \sigma_z$ is determined by

$$\sigma = \alpha - \gamma = 2E_L\varepsilon. \tag{S28}$$

Combining Eqs. (S27) and (S28), the incremental parameters in the Layer 1 are

$$\begin{aligned} \alpha_1 &= \mu_{L1} + \frac{2\mu_{T1} - \mu_{L1}}{E_{L1}}\sigma_1, \\ \gamma_1 &= \mu_{L1} + \frac{2\mu_{T1} - \mu_{L1} - E_{L1}}{E_{L1}}\sigma_1, \\ \beta_1 &= -\mu_{L1} + \frac{E_{L1}}{2} + \frac{\mu_{T1}}{2} + \frac{15\mu_{T1} - 10\mu_{L1} - 2E_{L1}}{2E_{L1}}\sigma_1, \end{aligned} \tag{S29}$$

where $\mu_{L1}$, $\mu_{T1}$, and $E_{L1}$ are transversely isotropic parameters of the Layer 1. The incremental parameters in the Layer 2 are

$$\begin{aligned} \alpha_2 &= \mu_{L2} + \frac{2\mu_{T2} - \mu_{L2}}{E_{L2}}\sigma_2, \\ \gamma_2 &= \mu_{L2} + \frac{2\mu_{T2} - \mu_{L2} - E_{L2}}{E_{L2}}\sigma_2, \\ \beta_2 &= -\mu_{L2} + \frac{E_{L2}}{2} + \frac{\mu_{T2}}{2} + \frac{15\mu_{T2} - 10\mu_{L2} - 2E_{L2}}{2E_{L2}}\sigma_2, \end{aligned} \tag{S30}$$

where $\mu_{L2}$, $\mu_{T2}$, and $E_{L2}$ are elastic parameters of the Layer 2. Inserting Eqs. (S29-S30) into Eqs. (S4-S8), the secular equation for the bilayer model with transversely isotropic material is obtained.

### S5.2 Parameter sensitivity study of the dispersion curve

Figure S5 show various numerical simulations of wave dispersion for different values of $\mu_{L1}$, $\mu_{T1}$, $E_{L1}$, $\mu_{L2}$, $\mu_{T2}$, and $E_{L2}$ for zero tensions (a-f) and finite tensions (g-h). The parameter $\mu_{L1}$ has a significant impact on high-frequency dispersion, while $\mu_{L2}$ notably affects low-frequency dispersion. The other parameters have a minimal influence on the dispersion. From the perspective of experimental fitting, the obtained shear moduli $\mu_1$ and $\mu_2$ are primarily consistent with the shear moduli of the lens capsule and cortex along the longitudinal (polar) direction. However, the fitted Young's moduli $E_1$ and $E_2$ may differ obviously from the real moduli $E_{L1}$ and $E_{L2}$.

When $\mu_{L1}/\mu_{T1}$ is increased from 1 (isotropic case) to 2, the phase velocities across 1 – 30 kHz is changed by less than 1% (Fig. S5g). When $E_{L1}/\mu_{L1}$ is increased from 3 (isotropic case) to 5, the phase velocities in 1 – 30 kHz is changed by ~ 8% (Fig. S5h). From the perspective of experimental fitting, since the degree of anisotropy of the lens is relatively weak [11], the error due to the assumption of isotropic lens should be minimal.

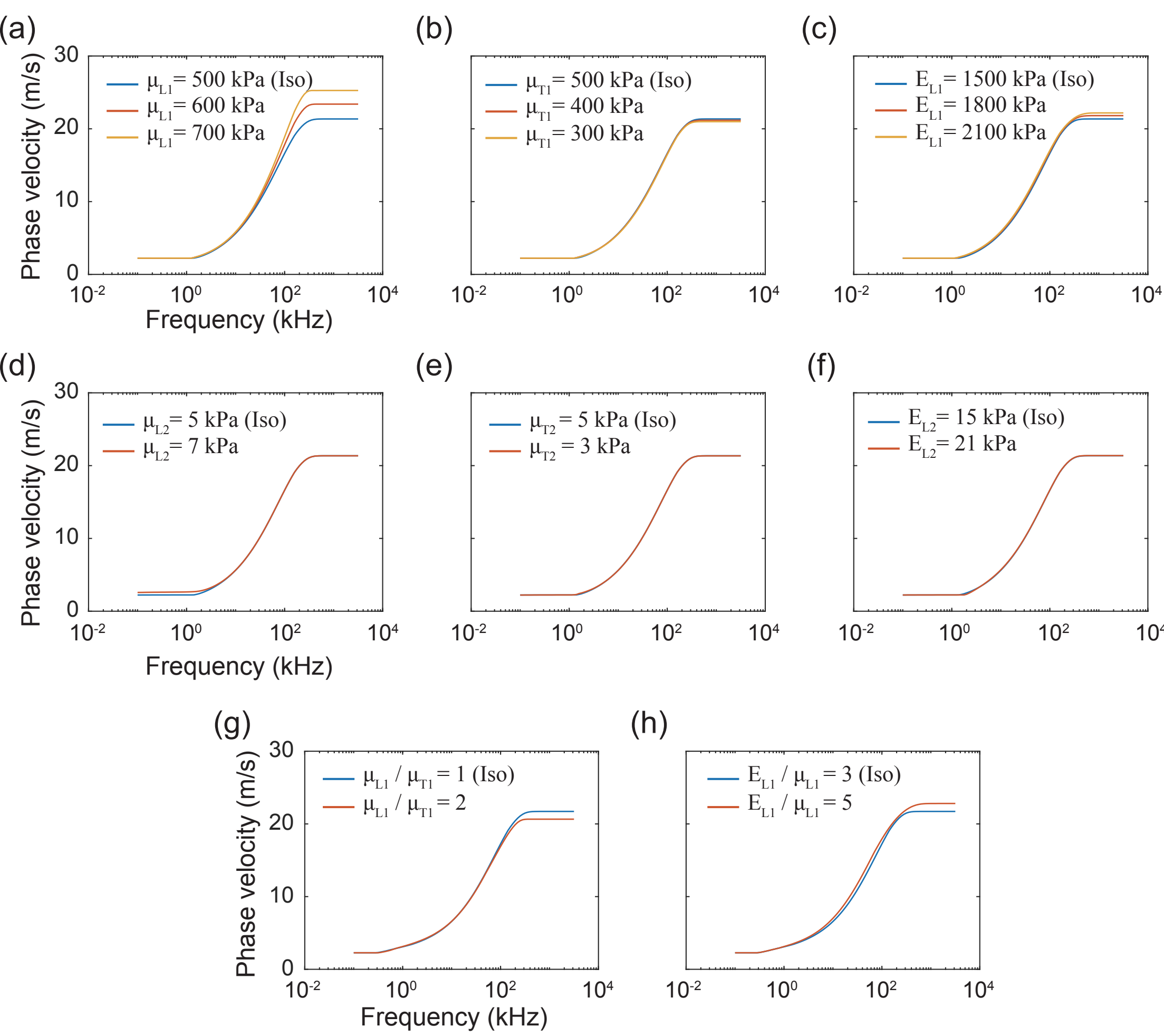

**Fig. S5**. Parameter sensitivity study of the dispersion curve. (**a-f**) Dispersion curves in the stress-free

state, with varying (a) $\mu_{L1}$, (b) $\mu_{T1}$, (c) $E_{L1}$, (d) $\mu_{L2}$, (e) $\mu_{T2}$, and (f) $E_{L2}$. Baseline parameters used were $\mu_{L1} = \mu_{T1} = 500$ kPa, $E_{L1} = 1500$ kPa, $\mu_{L2} = \mu_{T2} = 5$ kPa, $E_{L2} = 15$ kPa, $h = 55$ μm, and $\sigma_1 = \sigma_2 = 0$. (**g-h**) Dispersion curves in the stressed state, with varying (g) $\mu_{L1}/\mu_{T1}$ and (h) $E_{L1}/\mu_{L1}$. The same baseline parameters were used except for $\sigma_1 = 50$ kPa and $\sigma_2 = 0.5$ kPa.